\documentclass[prd,twocolumn,preprintnumbers,amsmath,amssymb,nofootinbib]{revtex4-2}

\usepackage[T1]{fontenc}
\usepackage[utf8]{inputenc}
\usepackage{microtype}
\usepackage{graphicx}
\usepackage{booktabs}
\usepackage{hyperref}
\usepackage{siunitx}
\usepackage{bm}
\usepackage{xcolor}
\usepackage{comment}

\AtBeginDocument{%
  \renewcommand\doi[1]{}%
  \renewcommand\url[1]{}%
}

\begin{document}

\title{Probing Planck-Scale Physics with High-Frequency Gravitational Waves}

\author{Stefano Profumo}
\affiliation{Department of Physics, University of California, Santa Cruz, CA 95064, USA}
\affiliation{Santa Cruz Institute for Particle Physics, University of California, Santa Cruz, CA 95064, USA}

\date{\today}


\begin{abstract}
\noindent We develop a framework for testing quantum gravity through the stochastic
gravitational-wave background produced by evaporating near-Planck-mass
primordial black holes. Because gravitons free-stream from the emission
region without rescattering, they preserve a direct spectral record of the
black-hole temperature--mass relation $T(M)$, a relation that is erased for all
other Hawking-radiated species by rapid thermalization. We translate six
representative phenomenological beyond-semiclassical frameworks (the generalized uncertainty
principle, loop quantum gravity, noncommutative geometry, asymptotic safety,
string/Hagedorn physics, and tunneling backreaction) into distinct $T(M)$
parametrizations and compute the resulting gravitational wave spectra
numerically. Modifications that suppress $T(M)$ shift the spectral peak by up
to ten decades in frequency, in some cases into the sensitivity bands of
next-generation interferometers or resonant-cavity detectors, while models
imposing a hard evaporation cutoff produce distinctive peak morphologies that
discriminate between quantum-gravity scenarios. We further discuss the impact of different choices for post-inflationary conditions in the very early universe.
We find that the relative spectral displacement between the standard Hawking
prediction and any modified model is cosmology-independent, hence spectral shape
rather than absolute peak frequency provides the cleanest probe of Planck-scale
physics.\end{abstract}

\maketitle

\section{Introduction}\label{sec:intro}
The marriage of quantum mechanics and general relativity remains one of the
most profound open challenges in theoretical physics~\cite{Carlip:2001wq, 
Kiefer:2007ria}. While quantum field theory in curved spacetime has provided
remarkable insights, most notably through Hawking's discovery that black holes
emit thermal radiation~\cite{Hawking:1974sw,Hawking:1975vcx} and the
associated thermodynamic interpretation of black-hole
mechanics~\cite{Bardeen:1973gs,Bekenstein:1973ur}, a complete, UV-finite
quantum theory of gravity remains elusive. The central difficulty is
experimental: the natural energy scale of quantum gravitational effects,
the Planck scale $E_{\rm Pl} \sim 10^{19}$\,GeV, lies some fifteen orders
of magnitude beyond the reach of any conceivable particle accelerator.

A number of indirect observational windows onto quantum gravity have
nevertheless been proposed and, in some cases, actively pursued.
Searches for Lorentz invariance violation exploit the fact that many
quantum-gravity frameworks predict energy-dependent modifications to
particle dispersion relations, which accumulate over cosmological distances
and can be constrained using high-energy photons from gamma-ray
bursts~\cite{AmelinoCamelia:1997gz,Xiao:2009xe} or TeV blazars observed
by imaging atmospheric Cherenkov telescopes~\cite{MAGIC:2020egb,
HESS:2011aa}.  Spacetime foam models predict a stochastic
distance fuzziness at the Planck scale that can in principle be probed
through phase decoherence in interferometers~\cite{Amelino-Camelia:1999xw,
Ng:2022sjc} or through the angular blurring of distant
point sources~\cite{Steinbring:2007db,Perlman:2014ila}.  In the
gravitational-wave sector, stochastic backgrounds produced during
inflation~\cite{Grishchuk:1974ny,Starobinsky:1979ty} or during a
putative bounce in loop quantum
cosmology~\cite{Bojowald:2001xe,Ashtekar:2006wn} encode
quantum-gravitational corrections to the primordial tensor spectrum that
future CMB polarization experiments may be able to
detect~\cite{CMB-S4:2020lpa}.  The generalized uncertainty principle (GUP) framework, originally motivated
by thought experiments in string theory and black-hole
physics~\cite{Maggiore:1993rv,Scardigli:1999jh}, predicts corrections to
atomic spectra, optomechanical oscillators, and scanning tunneling
microscopes at levels that are beginning to be
constrained~\cite{Pikovski:2011zk}.  Finally, the
observation of near-extremal black holes in X-ray
binaries~\cite{McClintock:2013vwa} and the direct imaging of black-hole
shadows by the Event Horizon Telescope~\cite{EventHorizonTelescope:2019dse,
EventHorizonTelescope:2022wkp} probe the strong-field geometry near
horizons and can, in principle, be used to search for quantum corrections
to the Kerr metric predicted by specific UV completions of
gravity~\cite{Carballo-Rubio:2022aed}.

Despite this breadth of activity, all existing probes are either limited
to Planck-suppressed corrections at energies far below $E_{\rm Pl}$ or to
qualitative signatures whose connection to a specific microscopic theory is
difficult to establish.  A direct probe of quantum-gravitational dynamics
operating \emph{at} the Planck scale, in a regime where the semiclassical
approximation genuinely breaks down, has remained out of reach.  The
universe itself may, however, provide a laboratory through relics from
the Planck epoch: primordial black holes (PBHs) formed in the early
universe~\cite{Carr:1974nx,Carr:2009jm,Carr:2020gox} with masses near the
Planck scale would have evaporated at temperatures $T \sim T_{\rm Pl}$,
precisely where quantum-gravitational corrections to the standard Hawking
law are expected to be largest.  As we argue in this paper, the
gravitational waves emitted during this process encode a detailed record
of those corrections, and that record may be within reach of a new
generation of high-frequency gravitational-wave detectors.


In 1974, Hawking demonstrated that black holes are not truly black but emit
thermal radiation with a characteristic temperature~\cite{Hawking:1974sw,
Hawking:1975vcx}
\begin{equation}
T_H = \frac{\hbar c^3}{8\pi G M k_B}
    \approx \frac{10^{-7}}{M/M_\odot} \, \mathrm{K},
\label{eq:hawking_temp}
\end{equation}
where $M$ is the black-hole mass.  This temperature arises from quantum
vacuum fluctuations near the event horizon and represents a profound
synthesis of quantum mechanics ($\hbar$), gravity ($G$), and thermodynamics
($k_B$), a result that has been re-derived in multiple independent frameworks
and is widely regarded as one of the most reliable predictions of
semiclassical gravity~\cite{Wald:1994yd,Birrell:1982ix}.  The corresponding
evaporation timescale is~\cite{Page:1976df}
\begin{equation}
\tau_{\rm ev} \sim \frac{M^3}{M_{\rm Pl}^2}\,t_{\rm Pl}
    \approx 10^{64} \left(\frac{M}{M_\odot}\right)^3 \mathrm{yr},
\end{equation}
where $M_{\rm Pl} = \sqrt{\hbar c/G} \approx 2.18\times10^{-5}$\,g is the
Planck mass and $t_{\rm Pl} = \sqrt{\hbar G/c^5} \approx 5.4\times10^{-44}$\,s
is the Planck time.  For a solar-mass black hole, therefore, the evaporation time vastly exceeds the
age of the universe,  a black hole of mass $M \sim 10^{15}$\,g evaporates
today, while one of mass $M \sim M_{\rm Pl}$ would have evaporated in a single
Planck time, a regime in which the derivation of Eq.~\eqref{eq:hawking_temp}
ceases to be trustworthy.

Hawking's derivation rests on the \emph{semiclassical approximation}: the
gravitational field is treated as a fixed classical background while matter
fields are quantized on top of it.  This approximation is well controlled
when the spacetime curvature at the horizon, $\mathcal{R} \sim M^{-2}$ in
Planck units, is small compared to the Planck scale.  As the black hole
evaporates, however, the horizon curvature grows without bound, and the
semiclassical treatment must eventually break down as $M \to M_{\rm Pl}$~\cite{Giddings:1992hh,Mathur:2009hf}.  

The nature of this breakdown is
one of the deepest unsolved problems in theoretical physics \cite{Brown:2024ajk, Lin:2025wof}.  Proposed
resolutions span a wide spectrum: the formation of stable or metastable
Planck-scale \emph{remnants} that halt evaporation and store the missing
information~\cite{Aharonov:1987tp,Chen:2015leg}; a unitary evaporation
process in which quantum information is gradually transferred to subtle
correlations in the outgoing radiation~\cite{Page:1993wv,Hawking:2005kf};
\emph{firewalls} or other departures from semiclassical physics at the
horizon that enforce unitarity at the cost of the equivalence principle
for infalling observers~\cite{Almheiri:2012rt}; topology-changing
transitions in which baby universes or wormholes encode the missing
degrees of freedom~\cite{Hawking:1987mz,Maldacena:2001kr}; and, most
recently, the discovery that including gravitational saddle-point
contributions (``islands'') in the Euclidean path integral reproduces a
unitary Page curve without invoking any exotic horizon
physics~\cite{Penington:2019npb,Almheiri:2019psf}.  While these proposals
differ sharply in their microscopic assumptions, they agree on the
qualitative conclusion that the canonical $T\propto M^{-1}$ relation in
Eq.~\eqref{eq:hawking_temp} cannot hold arbitrarily close to the Planck
scale.  If it did, black holes with $M\ll M_{\rm Pl}$ would reach temperatures
vastly exceeding $T_{\rm Pl}\sim10^{32}$\,K, entering a regime where
spacetime itself becomes quantum and the notion of a semiclassical geometry
loses meaning.  Various approaches to quantum gravity, including string
theory~\cite{Susskind:1993ws,Strominger:1996sh}, loop quantum
gravity (LQG)~\cite{Ashtekar:2004eh,Perez:2012wv}, asymptotic
safety~\cite{Reuter:2012id}, and the GUP~\cite{Adler:2001vs,Amelino-Camelia:2013sba}, all predict
modifications to black-hole thermodynamics in precisely this regime,
motivating the systematic phenomenological study we undertake below.


The quantum-gravity frameworks surveyed in Sec.~\ref{sec:qg_candidates}
share a common phenomenological implication: the standard semiclassical
relation $T \propto M^{-1}$ must be modified once black-hole evaporation
probes Planckian energies.  The nature of the correction depends on the
underlying theory, but the possibilities can be organized according to the
qualitative behavior of $T(M)$ as $M \to M_{\rm Pl}$.

In string-theoretic constructions, quantum and higher-derivative corrections
to the near-horizon geometry modify the thermodynamic relations in a
controlled way at small mass~\cite{Sen:2012dw,Castro:2012bc}, while
Hagedorn-density-of-states arguments suggest a transition to a string-dominated
phase with a maximal temperature once the string scale is reached.  In GUP
models, the deformed uncertainty relation leads to corrections that can be
organized as a systematic expansion in $M_{\rm Pl}/M$,
\begin{equation}
T(M) = \frac{1}{8\pi M}
\left[1 - \beta\left(\frac{M_{\rm Pl}}{M}\right)
      + \mathcal{O}\!\left(\frac{M_{\rm Pl}}{M}\right)^{\!2}\,\right],
\label{eq:TGUP_expansion}
\end{equation}
where $\beta$ is a dimensionless model-dependent parameter~\cite{Cavaglia:2003cz,Nozari:2005rq}.
Higher-order terms in this series can drive $T(M)$ to a maximum before
forcing it back to zero at a finite remnant mass, generically implying
altered entropy--area relations and long-lived Planck-scale
remnants~\cite{Adler:2001vs,Chen:2015leg,Carr:2015yqa}.  LQG- and
noncommutative-geometry--inspired solutions share this qualitative endpoint
behavior, while tunneling and backreaction corrections provide a
conservative, minimal-assumption realization of the same physics.

In each case the deviation from the Hawking law is confined to the final
stages of evaporation, leaving the well-tested (at least theoretically!) large-mass behavior intact.
In the following section we translate these theoretical possibilities into
a set of concrete, phenomenologically distinct $T(M)$ parametrizations
suitable for computing the gravitational-wave emission spectrum.


Among all particle species produced by Hawking evaporation, gravitons
occupy a privileged role as carriers of spectral information.  Every other
species, such as photons, neutrinos, and quarks, undergoes rapid thermalization in the primordial plasma once
emitted, erasing the imprint of the black-hole temperature on timescales
set by the interaction rate $\Gamma_{\rm int} \sim \alpha T^2/M_{\rm Pl}$,
where $\alpha$ is a representative coupling strength.  Thermalization is
therefore effective at all temperatures below $T \sim M_{\rm Pl}/\alpha$,
which for Standard Model couplings ($\alpha \lesssim 10^{-2}$) is at least
two orders of magnitude above the Planck temperature itself, comfortably
encompassing the entire evaporation history for the black-hole masses of
interest here.  Gravitons, interacting only gravitationally with strength
$\sim T/M_{\rm Pl} \ll 1$, free-stream from the point of emission without
further scattering~\cite{Kolb:1990vq,Weinberg:2008zzc}.  Their spectrum
therefore preserves a direct, undistorted record of $T(M)$ throughout the
evaporation, making the gravitational-wave background a uniquely clean probe
of near-Planckian black-hole thermodynamics.

For a black hole at temperature $T$, the power radiated in any particle
species of spin $s$ follows a Stefan--Boltzmann law weighted by
spin-dependent greybody factors that
account for the scattering and partial reflection of quanta by the curved
spacetime near the horizon~\cite{Page:1976df,MacGibbon:1991tj,Birrell:1982ix}.
The emission spectrum is thermal and strongly peaked at frequencies $f \sim
T(M)/h$, so the entire evaporation history maps onto a characteristic
frequency band in the emitted graviton spectrum.  For $M \sim M_{\rm Pl}$
the standard Hawking relation gives $T \sim T_{\rm Pl}$, implying emission
frequencies
\begin{equation}
f_{\rm em} \sim \frac{k_B T_{\rm Pl}}{h} \sim 10^{32}\,\mathrm{Hz}.
\end{equation}
After cosmological redshifting by a factor $(1+z_{\rm em})^{-1}$ set by
the epoch of evaporation, these ultra-high
frequencies map onto a present-day band potentially spanning the MHz--THz
regime, depending on the reheating temperature and expansion history, as we explore below in Sec.~\ref{sec:redshifting}.

A population of light PBHs with masses $M \sim 10^{-1}$--$10^2\,M_{\rm Pl}$,
formed in the early Universe and evaporating prior to Big Bang
Nucleosynthesis, would produce a stochastic gravitational-wave background
(SGWB) through their cumulative incoherent graviton
emission~\cite{Anantua:2008am,Papanikolaou:2020qtd}.  Unlike SGWBs from
compact binary coalescences or inflation, this background originates from
the quantum evaporation of horizons and probes physics at energies
approaching the Planck scale.  Its spectral shape is sensitive to $T(M)$, its
overall amplitude to the PBH abundance and mass distribution at formation,
and its peak frequency to the cosmological history between evaporation and
today, three largely independent handles that together make the SGWB a
powerful multi-parameter diagnostic of Planck-scale physics.


The direct detection of gravitational waves by LIGO and
Virgo~\cite{LIGOScientific:2016aoc,LIGOScientific:2018mvr} has established
gravitational-wave astronomy as a precision probe of the Universe, but
existing and planned detectors cover only a narrow slice of the accessible
frequency spectrum.  Ground-based interferometers (LIGO, Virgo, KAGRA, and
their third-generation successors ET and CE) are sensitive in the
$\sim10$--$10^4$\,Hz band~\cite{LIGOScientific:2014qfs}, the space-based
LISA mission will open the $\sim10^{-4}$--$10^{-1}$\,Hz window, and pulsar
timing arrays probe nHz frequencies~\cite{NANOGrav:2023gor}.  The regime
above $\sim10^5$\,Hz remains essentially unexplored, despite the fact that
a variety of well-motivated early-Universe processes, including first-order
phase transitions, cosmic string networks, and, as argued here, PBH
evaporation, are expected to produce signals in this band.

Bridging this gap has become an active area of research.  Proposed
high-frequency detection strategies exploit the coupling of gravitational
waves to electromagnetic fields or mechanical degrees of freedom, and
include: inverse-Gertsenshtein conversion of gravitational waves into
photons in strong magnetic fields~\cite{Gertsenshtein:1961,Cruise:2005ga,
Domcke:2020kcp}; resonant electromagnetic cavities that enhance this
conversion for narrowband signals~\cite{Chen:2016vkg,Berlin:2021txa}; and
optomechanical or levitated-sensor approaches sensitive to rapid spacetime
oscillations~\cite{Aggarwal:2021biu}.  While none of these has yet reached
astrophysically relevant sensitivity, they collectively demonstrate that the
MHz--GHz frontier is experimentally accessible in principle, and rapid
progress in quantum sensing and superconducting cavity technology is
narrowing the gap between current performance and the targets required to
probe early-Universe sources.

This frequency window is precisely where signals from near-Planck-mass PBH
evaporation are expected to appear after cosmological redshifting, making the high-frequency GW frontier a
natural, and, we argue, uniquely motivated, arena for testing quantum
gravity through an astronomical observable.


A detection of spectral features in the high-frequency SGWB inconsistent
with the standard Hawking prediction would constitute a qualitatively new
form of empirical access to quantum-gravitational physics.  The key
advantage of this probe over alternatives such as searches for Lorentz
invariance violation~\cite{AmelinoCamelia:2008qg,MAGIC:2020egb}, spacetime
foam--induced decoherence~\cite{Perlman:2014ila}, or
dispersion-relation modifications~\cite{Jacobson:2005bg} is interpretive
clarity: Hawking evaporation provides a framework in which the
large-mass behavior is firmly established semiclassically, deviations at
small mass can be systematically parameterized, and any inferred departure
from $T\propto M^{-1}$ points unambiguously to quantum-gravitational
dynamics at the horizon scale rather than to a generic violation of
low-energy effective field theory.

Connecting this signal to observation requires confronting the cosmological
redshifting of the emitted gravitons, which introduces both a challenge and
an opportunity.  The present-day spectrum depends not only on $T(M)$ but
also on the expansion history between evaporation and today, specifically,
the epoch at which evaporation completes, the reheating temperature and
entropy content of the radiation bath, the fractional PBH contribution to
the total energy density, and the possible presence of nonstandard expansion
phases such as early matter domination or kination.  Different cosmological
scenarios can shift the spectral peak by many decades in frequency and alter
the overall amplitude by orders of magnitude.  Rather than treating this
as an irreducible systematic, however, one can exploit the fact that the
spectral shape, peak frequency, and amplitude encode \emph{correlated}
information about both $T(M)$ and the cosmological environment: the
redshift problem, properly framed, becomes a source of additional diagnostic
power.  


The remainder of this paper is organized as follows.
Section~\ref{sec:qg_candidates} surveys the quantum-gravity frameworks
that motivate modifications to the Hawking temperature--mass relation,
including the GUP, noncommutative geometry,
loop quantum gravity, asymptotic safety, string theory, and
tunneling/backreaction corrections.
Section~\ref{sec:modified} translates these motivations into a set of
concrete, phenomenologically distinct $T(M)$ parametrizations and derives
their qualitative consequences for the evaporation history and graviton
emission spectrum.
Section~\ref{sec:gw_production} describes the production of a stochastic
gravitational-wave background from light PBHs, establishes why the signal
is stochastic in nature, and presents the integral formula connecting the
instantaneous graviton spectrum to the present-day $\Omega_{\rm GW}(f_0)$ (see eq.~\eqref{eq:OmegaGW_def} for a definition thereof).
Section~\ref{sec:redshifting} treats the cosmological redshifting problem,
covering the standard radiation-dominated mapping, the realistic upper bound
set by the reheating temperature, and the modifications arising from
early matter domination, kination, and nonstandard reheating histories.
Section~\ref{sec:comparison} presents the numerical GW spectra for all six
modified-temperature models across a range of parameters and initial masses,
compares them with detector sensitivity envelopes, and discusses the
experimental requirements for resonant-cavity detection.
Section~\ref{sec:conclusions} summarizes our conclusions and outlines
directions for future work. Throughout, we retain all fundamental constants
($\hbar$, $c$, $G$, $k_B$) explicitly in dimensional
expressions; the numerical spectra in
Sec.~\ref{sec:comparison} are computed in the
dimensionless units defined in
Appendix~\ref{sec:app}, with all masses measured
in units of~$M_{\rm Pl}$.


\section{Quantum-Gravity Candidates Affecting Evaporation Near $M_{\rm Pl}$}
\label{sec:qg_candidates}


As the black-hole mass approaches the Planck scale, curvature invariants
at the horizon become large, quantum fluctuations of the geometry are no
longer parametrically suppressed, and the notion of a fixed classical
background ceases to be reliable.  Any complete theory of quantum gravity
must specify how the temperature--mass relation $T(M)$ is modified in this
regime.  While no fully predictive theory is currently available, a number
of well-studied frameworks provide qualitative guidance, and their
qualitative predictions for $T(M)$ near $M_{\rm Pl}$ are sufficiently
distinct to motivate a systematic phenomenological survey.  We outline each
in turn.

\subsection{Generalized Uncertainty Principle}

The GUP incorporates a minimal
measurable length into the structure of quantum mechanics by deforming
the Heisenberg relation to include Planck-suppressed
corrections~\cite{Adler:2001vs,Scardigli:1999jh}, leading to altered
dispersion relations and modified phase-space structure at high energies.
Applied to black-hole thermodynamics, these deformations modify the
relation between horizon size and characteristic emission wavelength,
generically driving $T(M)$ to a maximum before forcing it back toward
zero as $M$ approaches a minimum value of order $M_{\rm Pl}$~\cite{Cavaglia:2003cz,Nozari:2005rq}.  This behavior regulates the
semiclassical runaway and is commonly interpreted as signaling the
formation of a stable or metastable Planck-scale remnant~\cite{Adler:2001vs,Chen:2015leg}.

\subsection{Noncommutative Geometry}

In noncommutative-geometry--inspired models, spacetime coordinates satisfy
nontrivial commutation relations, effectively smearing point-like sources
over a minimal length scale~\cite{Nicolini:2006nc}.  Black-hole solutions
constructed in this framework replace the classical singularity with a
regular core, and the horizon structure is modified at small radius.  The
resulting Hawking temperature reaches a finite maximum and then decreases
to zero at a minimal mass~\cite{Nicolini:2005vd}, qualitatively similar to
the GUP outcome but with a temperature suppression that is quadratic in
$M_{\rm min}/M$ rather than the square-root suppression characteristic of
many LQG constructions (see below).  This difference in the functional form
of $T(M)$ near $M_{\rm min}$ is directly reflected in the width and
asymmetry of the resulting gravitational-wave spectrum.

\subsection{Loop Quantum Gravity and Polymerized Black Holes}

Loop quantum gravity  provides a nonperturbative,
background-independent quantization of spacetime geometry in which areas
and volumes acquire discrete spectra.  Effective black-hole solutions
inspired by LQG replace the classical singularity with a quantum region
governed by polymerized geometry~\cite{Modesto:2006LQG,Ashtekar:2005qt},
and many such constructions predict a transition to a ``Planck star''---a
compact, Planck-density object whose evaporation history differs markedly
from the standard picture~\cite{Rovelli:2014PlanckStar}.  The temperature
either vanishes smoothly or remains finite as $M \to M_{\rm min}$,
suppressing the Hawking runaway and producing an extended or halted
evaporation phase whose GW signature we compute in Sec.~\ref{sec:modified}.

\subsection{Asymptotic Safety and Renormalization-Group Improvements}

In the asymptotic safety scenario, gravity possesses a nontrivial
ultraviolet fixed point that renders Newton's constant scale-dependent at
high energies~\cite{Weinberg:1979as,Reuter:1998rg}.
Renormalization-group--improved black-hole solutions derived in this
framework exhibit a running $G(k) \sim G_0/(1 + \xi G_0 k^2)$ near the
horizon, where $k \sim M^{-1}$ is the relevant momentum scale~\cite{Bonanno:2000AS}.  The resulting $T(M)$ relation deviates from the
semiclassical form once the running becomes important, typically approaching
a finite maximum temperature and yielding a modified evaporation endpoint.
Unlike the GUP and NC scenarios, asymptotic safety does not introduce a
minimal mass per se, but rather a smooth infrared--ultraviolet interpolation
that softens the late-time luminosity.

\subsection{String-Theoretic Considerations and the Hagedorn Temperature}

String theory provides a UV-complete framework in which black holes
transition into highly excited string states once their mass approaches
the string scale~\cite{Horowitz:1997string}.  The exponentially growing
density of string states implies a maximal (Hagedorn) temperature $T_H$
at which thermodynamic systems absorb energy without further heating,
suggesting that black-hole evaporation may be regulated by a smooth
crossover from a black-hole phase to a string-dominated
phase~\cite{Amati:1989Strings,Atick:1988si}.  Unlike the remnant-forming
scenarios above, this predicts a \emph{limiting} rather than a vanishing
temperature: $T(M) \to T_H$ as $M \to 0$, with $T_H$ set by the string
tension.  The resulting graviton spectrum has a qualitatively distinct
high-frequency cutoff whose location encodes the string scale directly.

\subsection{Tunneling and Backreaction Corrections}

A conceptually distinct modification arises from enforcing energy
conservation during the emission process.  In the original semiclassical
derivation, Hawking radiation is computed on a fixed background, ignoring
the fact that each emitted quantum reduces the black-hole mass and thereby
modifies the geometry seen by subsequent quanta.  Interpreting Hawking
radiation as a tunneling process across the horizon~\cite{Parikh:2000dx}
enforces this energy conservation explicitly: the emission of a quantum of
energy $\omega$ shifts the black-hole mass from $M$ to $M-\omega$, and
the corrected tunneling rate
\begin{equation}
\Gamma(\omega) \propto \exp\!\bigl[\Delta S_{\rm BH}(M-\omega)
                                  - \Delta S_{\rm BH}(M)\bigr]
\end{equation}
deviates from exact thermality at order $\omega/M$.  To leading order,
these backreaction corrections reduce the effective temperature relative
to the Hawking value~\cite{Parikh:2004,Medved:2002}, providing a
conservative lower bound on the magnitude of Planck-scale modifications
that does not invoke any new microscopic degrees of freedom.  In this
sense, tunneling corrections serve as a minimal benchmark against which
the more radical modifications of the preceding subsections can be compared.

\subsection{Shared Features and Phenomenological Organization}

Despite their different microscopic origins, the frameworks above share
three qualitative features that are directly relevant for gravitational-wave
phenomenology.  First, all introduce a characteristic mass scale---a
minimal length, a running coupling scale, a string tension, or a remnant
mass---that regulates the semiclassical divergence of $T(M)$ as
$M \to M_{\rm Pl}$.  Second, the nature of this regulation falls into two
broad classes: \emph{limiting-temperature} models (plateau and Hagedorn),
in which $T$ saturates at a finite value, and \emph{cooling} models (GUP,
NC, LQG, tunneling), in which $T$ reaches a maximum and subsequently
decreases.  Third, the corrections to other thermodynamic quantities---
entropy, heat capacity, and luminosity---follow from the same physics and
can significantly prolong or arrest the final evaporation phase, with direct
consequences for the duration and spectral content of graviton emission.
These shared features motivate the phenomenological parametrizations of
$T(M)$ introduced in Sec.~\ref{sec:modified}, which are designed to
capture the essential physics of each class in a form amenable to
quantitative gravitational-wave calculations.


\section{Modified Black-Hole Temperature--Mass Relations}
\label{sec:modified}


\subsection{Standard Hawking law}
\label{subsec:hawking_TM}

In semiclassical gravity, a stationary black hole radiates thermally at the
Hawking temperature determined by its surface gravity $\kappa$,
\begin{equation}
T_{\rm H} \;=\; \frac{\hbar\,\kappa}{2\pi k_{\rm B} c}\,,
\label{eq:TH_kappa}
\end{equation}
a relation that holds for generic horizons and underlies the identification
of black holes as thermodynamic systems~\cite{Hawking:1975vcx,Bardeen:1973gs,
Wald:1994yd}.  For a four-dimensional, asymptotically flat Schwarzschild
black hole,
\begin{equation}
\kappa \;=\; \frac{c^4}{4GM}\,,\qquad
T_{\rm H}(M)\;=\;\frac{\hbar c^3}{8\pi G k_{\rm B} M}
\;\equiv\; \frac{1}{8\pi}\,\frac{M_{\rm Pl}^2 c^2}{k_{\rm B} M}\,,
\label{eq:TH_Schw}
\end{equation}
so that $T_{\rm H}\propto M^{-1}$.  In convenient numerical form,
\begin{equation}
T_{\rm H}\simeq 6.17\times 10^{-8}\,\mathrm{K}\,\left(\frac{M_\odot}{M}\right)
\simeq 1.06\,\mathrm{GeV}\,\left(\frac{10^{13}\,\mathrm{g}}{M}\right)\,,
\label{eq:TH_numeric}
\end{equation}
where the second expression is particularly useful for primordial black holes.
This scaling implies that as a black hole evaporates its temperature rises
monotonically, driving an increasingly rapid late-time evolution.

The corresponding semiclassical mass-loss rate is
\begin{equation}
\frac{\mathrm{d}M}{\mathrm{d}t}\;=\;-\frac{\alpha(M)}{M^{2}}\,,
\label{eq:dMdt_std}
\end{equation}
where $\alpha(M)$ encodes the effective number of emitted species and the
relevant greybody factors; in the high-temperature limit where all Standard
Model degrees of freedom are accessible, $\alpha(M)$ is approximately
constant between threshold crossings~\cite{Page:1976df,MacGibbon:1991tj}.
Integrating Eq.~\eqref{eq:dMdt_std} gives
\begin{equation}
\tau_{\rm evap}(M_0)\;\simeq\;\frac{M_0^3}{3\,\alpha}\;\propto\; M_0^3\,,
\label{eq:tau_M3}
\end{equation}
which in conventional units yields
\begin{equation}
\tau_{\rm evap}\;\simeq\; 4\times 10^{-28}\,\mathrm{s}\,
\left(\frac{M_0}{1\,\mathrm{g}}\right)^3
\;\simeq\; 4\times 10^{17}\,\mathrm{s}\,
\left(\frac{M_0}{5\times 10^{14}\,\mathrm{g}}\right)^3
\label{eq:tau_numeric}
\end{equation}
up to $\mathcal{O}(1)$ factors~\cite{MacGibbon:1991tj,Carr:2009jm}.  The
instantaneous emission spectrum is approximately thermal at temperature
$T_{\rm H}(M)$, with spin-dependent greybody factors encoding propagation
through the curved spacetime outside the horizon~\cite{Page:1976df}.
Graviton emission is strongly suppressed at large mass but becomes
appreciable in the final stages as the temperature rises, making the
high-frequency tail of the SGWB directly sensitive to the late-time
behavior of $T(M)$~\cite{Page:1976df,Anantua:2008am}.

An additional consideration is black-hole rotation.  For Kerr black holes
the Hawking temperature is reduced relative to the Schwarzschild value at
fixed mass, while the greybody factors become strongly spin- and
mode-dependent.  Bosonic radiation is further enhanced through superradiant
amplification for modes satisfying $\omega < m\,\Omega_{\rm H}$, where
$\Omega_{\rm H}$ is the horizon angular velocity~\cite{Starobinsky:1973sq,
Teukolsky:1974yv,Page:1976ki}, so rapidly spinning black holes radiate a
substantially larger fraction of their power in high-spin channels,
including gravitons, and spin down efficiently during
evaporation~\cite{Page:1976ki,Page:1977um}.  For primordial black holes,
whose initial spin distribution is model dependent, rotation can therefore
significantly affect both the amplitude and spectral shape of the
high-frequency SGWB~\cite{Carr:2009jm,Dong:2015yjs}.  We defer a systematic
treatment of the Kerr case to future work and focus here on the
non-rotating limit, which captures the essential physics of the
temperature--mass modifications of interest.

Equation~\eqref{eq:TH_Schw} provides the baseline against which any
Planck-scale modification is defined.  In the remainder of this section
we introduce representative deformations that capture qualitatively distinct
possibilities for how the semiclassical evaporation law may be regulated as
$M\to M_{\rm Pl}$.

\subsection{Plateau and Hagedorn-like models}
\label{subsec:plateau_hagedorn}

A common expectation in UV-complete or UV-regulated frameworks is that the
divergent rise of the Hawking temperature is softened near the fundamental
scale.  In string theory the exponentially growing density of states allows
thermodynamic systems to approach a limiting (Hagedorn) temperature,
suggesting a crossover from a black-hole phase to a string-dominated
regime~\cite{Horowitz:1997string,Amati:1989Strings}.  Motivated by this
behavior, we model the approach to a maximal temperature through a smooth
saturation of $T(M)$:
\begin{equation}
T_{\rm plat}(M)\;=\;
\frac{1}{8\pi}\frac{M_{\rm Pl}^2}{\sqrt{M^2+M_c^2}}\,,
\label{eq:T_plateau}
\end{equation}
where $M_c$ is a characteristic crossover mass.  For $M\gg M_c$,
Eq.~\eqref{eq:T_plateau} reduces to the standard Hawking law; for
$M\ll M_c$ it asymptotes to a finite maximal temperature
\begin{equation}
T_{\rm max}\;=\;\frac{1}{8\pi}\frac{M_{\rm Pl}^2}{M_c}\,.
\label{eq:Tmax_plateau}
\end{equation}
A related parametrization that more directly mimics a Hagedorn-like limiting
temperature is
\begin{equation}
T_{\rm Hag}(M)\;=\;\frac{T_H}{1+\left(\dfrac{M}{M_H}\right)^{\!p}}\,,
\qquad
T_H \equiv \frac{1}{8\pi}\frac{M_{\rm Pl}^2}{M_H}\,,
\label{eq:T_Hagedorn}
\end{equation}
with $p>0$ controlling the sharpness of the crossover; for $M\gg M_H$ this
reproduces $T\propto M^{-1}$, while for $M\ll M_H$ the temperature saturates
to $T_H$~\cite{Amati:1989Strings,Atick:1988si}.  Both
Eqs.~\eqref{eq:T_plateau} and~\eqref{eq:T_Hagedorn} should be regarded as
effective descriptions of UV-regulated evaporation rather than first-principles
predictions; the crossover scales $M_c$ and $M_H$ encode the onset of new
degrees of freedom and control the location of spectral features in the
graviton background.

The effect on the evaporation history is immediate.  The standard Hawking
luminosity $L\propto M^{-2}$ implies a runaway at late times.  Once $T$
saturates in a plateau/Hagedorn model, however, $L\propto M^2 T_{\rm max}^4$
and the evaporation rate slows dramatically.  Absent an additional
termination mechanism, this prolonged low-luminosity phase can produce
quasi-stable Planck-scale relics whose lifetimes are parametrically enhanced
relative to the semiclassical expectation~\cite{Chen:2015leg,Carr:2015yqa}.

\subsection{Cooling models and post-peak temperature decrease}
\label{subsec:cooling_TM}

A qualitatively distinct possibility is that $T(M)$ reaches a maximum at a
finite mass and subsequently \emph{decreases}, so that the late stages of
evaporation are characterized by progressively softer emission.  A
convenient parametrization is
\begin{equation}
T_{\rm cool}(M)
\;=\;
\frac{1}{8\pi}\,
\frac{M_{\rm Pl}^2\,M}{M^2 + M_c^2}\,,
\label{eq:T_cooling}
\end{equation}
where $M_c$ sets the peak-temperature mass.  For $M\gg M_c$,
Eq.~\eqref{eq:T_cooling} reduces to $T\propto M^{-1}$; for $M\ll M_c$ it
gives $T\propto M$, so the temperature vanishes linearly as $M\to 0$.  The
peak occurs at $M=M_c$ with
\begin{equation}
T_{\rm max}
\;=\;
\frac{1}{16\pi}\,
\frac{M_{\rm Pl}^2}{M_c}\,.
\end{equation}
This qualitative behavior---a maximum followed by a decrease---arises
naturally in GUP frameworks, noncommutative geometry, and LQG-inspired
constructions (as discussed in Sec.~\ref{sec:qg_candidates}), making
Eq.~\eqref{eq:T_cooling} a useful phenomenological representative of that
entire class.

The dynamical consequences are substantial: for $M\ll M_c$ the luminosity
scales as $L\propto M^6$, so the evaporation rate slows dramatically and the
black hole either asymptotes to a long-lived remnant or undergoes an extended
phase of low-temperature emission~\cite{Chen:2015leg,Carr:2015yqa}.  For
gravitational-wave phenomenology, because the hardest graviton emission occurs
at intermediate masses $M\sim M_c$, the SGWB in cooling scenarios exhibits a
pronounced peak at the frequency set by $T_{\rm max}$, followed by a rapid
high-frequency falloff---a spectral shape qualitatively distinct from both the
standard Hawking case and the plateau/Hagedorn models.

\subsection{GUP-corrected, noncommutative, and LQG-inspired forms}
\label{subsec:GUP_NC_LQG}

The frameworks reviewed in Sec.~\ref{sec:qg_candidates} motivate a class of
$T(M)$ relations that share the feature of a minimum mass $M_{\rm min}$ below
which evaporation ceases.  The key phenomenological distinction among them is
the functional form of $T(M)$ as $M\to M_{\rm min}^+$, which controls the
width and asymmetry of the resulting GW spectral peak.

For GUP-based models, the correction to the Hawking temperature can be written
schematically as
\begin{equation}
T_{\rm GUP}(M)
\;=\;
\frac{1}{8\pi M}
\left[
1 - \mathcal{O}\!\left(\frac{M_{\rm Pl}}{M}\right)^2
\right],
\end{equation}
with higher-order terms driving $T$ to zero at a finite remnant
mass~\cite{Cavaglia:2003cz,Nozari:2005rq}.  A cleaner parametrization that
captures this behavior exactly and enforces a minimal mass is
\begin{equation}
T_{\rm GUP}(M) \;=\;
\frac{A}{2M/\beta}\left[1 - \sqrt{1 - \frac{\beta}{M^2}}\right],
\qquad M\geq\sqrt{\beta}\,,
\label{eq:TGUP_exact}
\end{equation}
where $\beta>0$ is the deformation parameter and $\sqrt{\beta}$ plays the
role of the remnant mass.

The LQG-inspired and noncommutative-geometry forms are distinguished by the
power with which $T$ vanishes near $M_{\rm min}$:
\begin{align}
T_{\rm LQG}(M) &\;=\; \frac{A}{M}
\sqrt{1-\left(\frac{M_{\rm min}}{M}\right)^{\!2}}\,,
\qquad M\geq M_{\rm min}\,,
\label{eq:TLQG}\\[4pt]
T_{\rm NC}(M) &\;=\; \frac{A}{M}
\left[1-\left(\frac{M_{\rm min}}{M}\right)^{\!2}\right],
\qquad M\geq M_{\rm min}\,.
\label{eq:TNC}
\end{align}
In both cases $T\to 0$ as $M\to M_{\rm min}$, but the LQG suppression is
$\propto(M-M_{\rm min})^{1/2}$ while the NC suppression is
$\propto(M-M_{\rm min})$, a quadratic rather than square-root vanishing.
This difference is directly imprinted on the GW spectrum: the NC spectral
peak is narrower and more sharply defined than its LQG counterpart at
comparable $M_{\rm min}$, providing a potential morphological discriminant
between the two frameworks.

\subsection{Tunneling and backreaction-based corrections}
\label{subsec:tunneling_backreaction}

Interpreting Hawking radiation as a quantum tunneling process across the
horizon~\cite{Parikh:2000dx} enforces energy conservation explicitly: the
emission of a quantum of energy $\omega$ shifts the black-hole mass from $M$
to $M-\omega$, modifying the geometry seen by the tunneling particle.  The
corrected emission rate is
\begin{equation}
\Gamma(\omega) \;\propto\;
\exp\!\bigl[\Delta S_{\rm BH}(M-\omega) - \Delta S_{\rm BH}(M)\bigr]\,,
\end{equation}
which deviates from exact thermality at order $\omega/M$ and can be recast
as an effective temperature suppression,
\begin{equation}
T_{\rm eff}(M) \;=\; T_{\rm H}(M)
\left[1 - \mathcal{O}\!\left(\frac{T_{\rm H}}{M}\right)\right].
\end{equation}
These corrections are formally suppressed by $M_{\rm Pl}/M$ but become
$\mathcal{O}(1)$ when $M\sim M_{\rm Pl}$~\cite{Parikh:2004,Medved:2002},
producing a softened and prolonged final evaporation phase that is
qualitatively similar to the cooling scenarios above.  Unlike GUP or NC
models, tunneling corrections require no new microscopic degrees of freedom
and in this sense provide a \emph{minimal} benchmark for Planck-scale
modifications of $T(M)$.

\subsection{Comparative summary and physical interpretation}
\label{subsec:comparative_summary}

The models introduced above can be organized into two phenomenologically
distinct classes according to their behavior as $M\to 0$ (or
$M\to M_{\rm min}$):

\emph{Limiting-temperature models} (plateau, Hagedorn): $T(M)$ saturates at
a finite value $T_{\rm max}$ set by a new fundamental scale.  The evaporation
rate slows at late times, redistributing power toward lower frequencies
relative to standard Hawking.

\emph{Cooling models} (GUP, NC, LQG, tunneling): $T(M)$ reaches a maximum
at an intermediate mass and then decreases toward zero, concentrating
graviton emission in a finite frequency window and producing a sharply
peaked $\Omega_{\rm GW}(f)$ with a steep high-frequency cutoff.

\begin{figure}[t]
\centering
\includegraphics[width=0.95\linewidth]{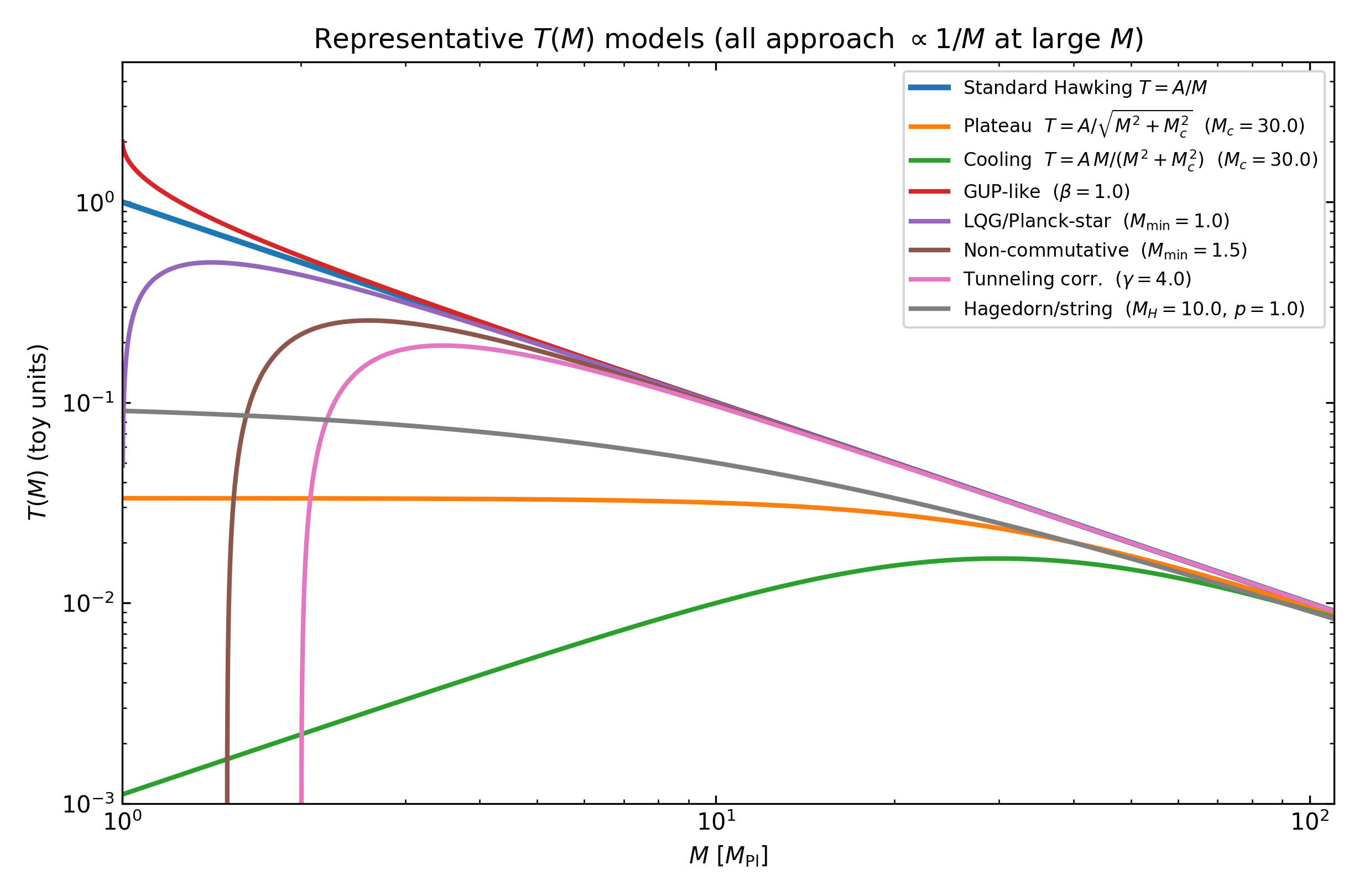}
\caption{Representative temperature--mass relations $T(M)$ for the modified
black-hole evaporation models considered in this work, shown on logarithmic
axes.  All curves asymptote to the standard Hawking behavior $T\propto M^{-1}$
at large mass while exhibiting qualitatively distinct behavior near the Planck
scale: plateau/Hagedorn-like saturation, cooling (post-peak decrease), GUP-,
LQG-, and noncommutative-geometry--inspired forms with minimum masses, and
tunneling/backreaction corrections.}
\label{fig:TM_models_log}
\end{figure}

\begin{figure}[t]
\centering
\includegraphics[width=0.95\linewidth]{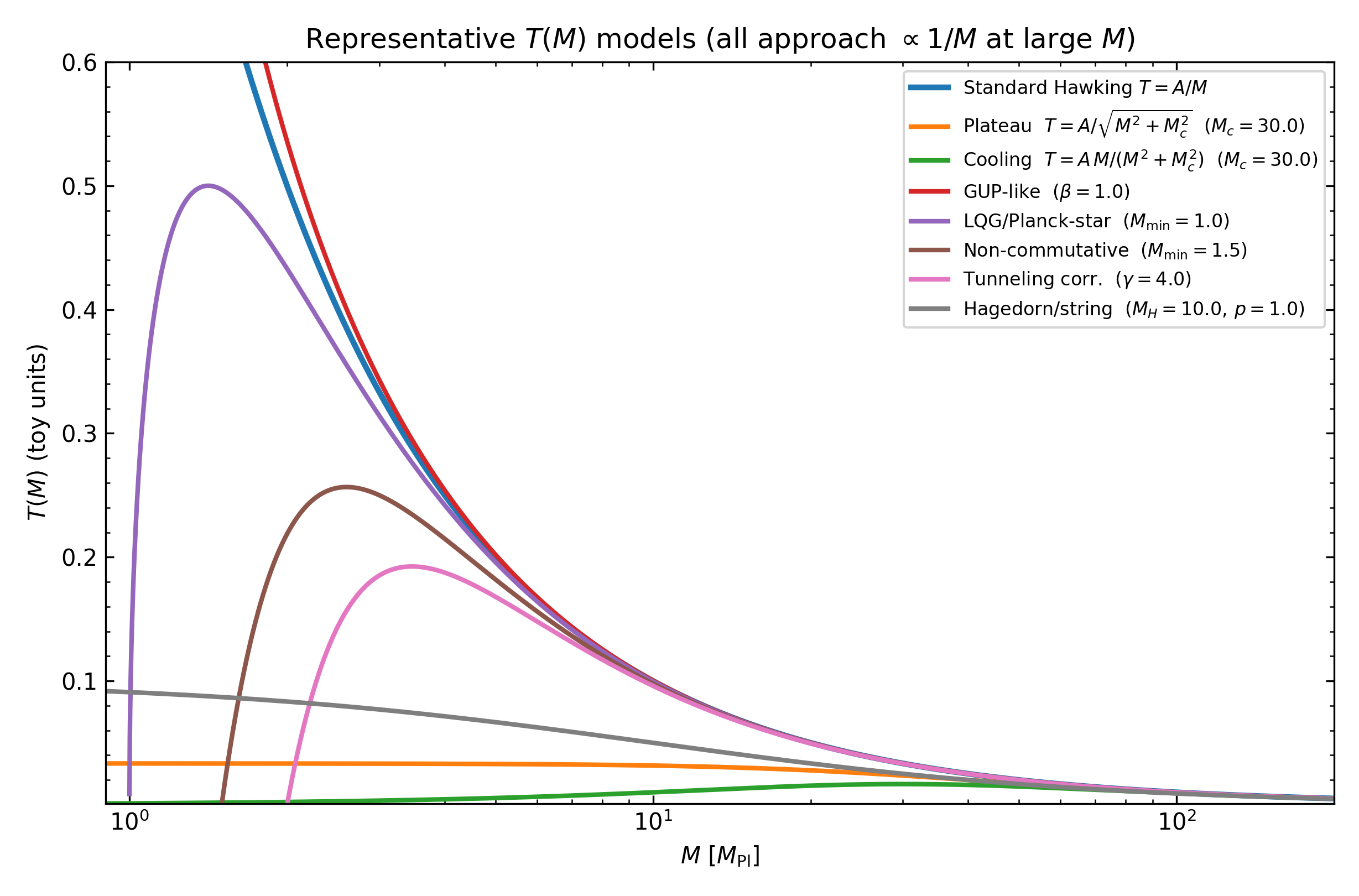}
\caption{Same as Fig.~\ref{fig:TM_models_log} on linear axes, highlighting
the relative ordering of peak temperatures, the presence or absence of maxima,
and the suppression or termination of evaporation near a minimum mass.}
\label{fig:TM_models_lin}
\end{figure}

Within the cooling class, the sharpness of the spectral peak is further
controlled by the functional form of $T$ near $M_{\rm min}$: the NC
quadratic suppression yields a narrower peak than the LQG square-root
suppression, and the tunneling correction---the most conservative of the
group---yields the broadest.  These differences, summarized in
Figs.~\ref{fig:TM_models_log} and \ref{fig:TM_models_lin}, translate directly
into distinctive spectral morphologies in $\Omega_{\rm GW}(f_0)$ that we
compute in Sec.~\ref{sec:comparison}. The logarithmic representation emphasizes the shared
semiclassical asymptotics and makes clear that deviations from the Hawking law
are confined to a narrow mass interval near $M_{\rm Pl}$.  The linear-scale
plot highlights the qualitative distinctions---peak heights, the onset of
cooling, and the presence of a minimum mass---that translate into the spectral
features discussed in what follows (Sec.~\ref{sec:comparison} in particular).

\begin{figure}[t]
\centering
\includegraphics[width=0.95\linewidth]{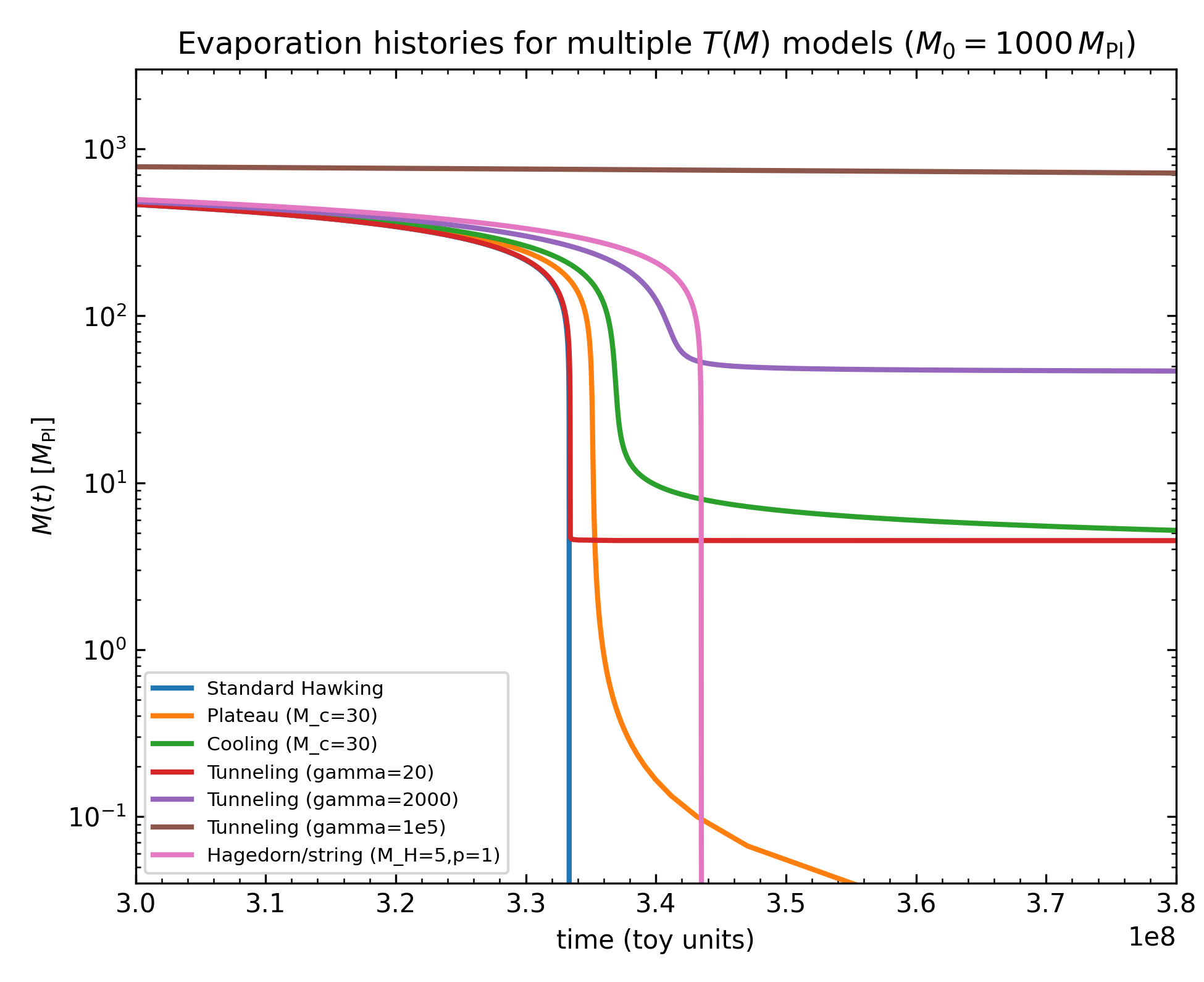}
\caption{Illustrative evaporation histories $M(t)$ for initial mass
$M_0=10^3\,M_{\rm Pl}$ in several modified $T(M)$ scenarios.  The figure shows how modified models
can (i) delay the late-stage evolution, (ii) produce a prolonged low-mass
tail, or (iii) effectively arrest evaporation at a finite remnant mass.
Time is shown in the dimensionless units of Appendix~\ref{sec:app}.}
\label{fig:Mt_models}
\end{figure}

Figure~\ref{fig:Mt_models} illustrates how the $T(M)$ classes map onto
evaporation histories $M(t)$.  In the standard Hawking picture, most of the
lifetime is spent near $M_0$, followed by a brief terminal plunge as the
temperature diverges.  Plateau/Hagedorn models stretch this final phase by
suppressing the luminosity growth; cooling models extend it further by
allowing the temperature---and hence the power---to decrease at late times,
producing an extended low-mass tail.  Models where $T\to 0$ at a finite
$M_{\rm min}$ produce trajectories that asymptote to a nonzero mass,
mimicking remnant formation on cosmological timescales.  These qualitative
differences in $M(t)$ map directly onto the emission history and hence the
present-day GW spectrum: a rapid plunge concentrates power at the highest
frequencies, while a stalled or prolonged tail redistributes it toward lower
frequencies and reduces the peak amplitude,

\section{Gravitational-Wave Production from Light PBHs}
\label{sec:gw_production}

Primordial black holes with masses far below the astrophysical scale produce
gravitational radiation through several distinct mechanisms during their
evaporation.  Because PBHs form with a broad spatial distribution, evaporate
independently, and emit Hawking quanta with random phases, their cumulative
GW output is inherently incoherent, giving rise to a SGWB rather than a resolvable transient
signal.  In this section we describe the relevant emission processes,
establish the stochastic nature of the signal, and present the integral
formula connecting the evaporation physics to the observable spectral energy
density $\Omega_{\rm GW}(f_0)$.

\subsection{Emission mechanisms}
\label{subsec:emission_mechanisms}

\paragraph{Direct graviton emission.}
In the semiclassical description of Hawking evaporation, black holes emit all
particle species with masses below the instantaneous temperature, including
gravitons~\cite{Page:1976df,MacGibbon:1991tj}.  Graviton emission is
suppressed relative to lower-spin species at low temperatures due to the
higher angular-momentum barrier in the spin-2 greybody factors, but becomes
an appreciable fraction of the total radiated power as the temperature
approaches $M_{\rm Pl}$~\cite{Page:1976df}.  Because gravitons interact only
gravitationally, they free-stream from the horizon without further scattering
(as established in Sec.~\ref{sec:intro}), preserving a direct record of
$T(M)$ in the emitted spectrum.

\paragraph{Secondary gravitational-wave production.}
GWs can also be generated indirectly through the decay, annihilation, or
hadronization of energetic particles produced alongside the direct Hawking
quanta~\cite{Anantua:2008am}.  Relativistic particle
cascades and the decay of heavy intermediates can source tensor perturbations
in the surrounding plasma.  These secondary processes are typically
subdominant to direct graviton emission at the highest frequencies, but can
broaden the GW spectrum and introduce model-dependent features tied to the
particle content of the theory. We neglect this emission mechanism in what follows.

\paragraph{Evaporation-induced cosmological dynamics.}
If PBHs constitute a non-negligible fraction of the total energy density
prior to evaporation, their disappearance injects entropy and radiation into
the primordial plasma, modifying the subsequent expansion
history~\cite{Carr:2009jm,Inomata:2019zqy}.  This energy injection does not
directly source coherent GWs, but it alters the redshifting and dilution of
previously emitted gravitons, making the final observed spectrum sensitive to
the cosmological environment at the time of evaporation.  We treat this
dependence in detail in Sec.~\ref{sec:redshifting}.

\subsection{Stochastic nature and spectral characterization}
\label{subsec:stochastic}

The observable quantity is the spectral energy density
\begin{equation}
\Omega_{\rm GW}(f) \;\equiv\; \frac{1}{\rho_c}\,\frac{{\rm d}\rho_{\rm GW}}{{\rm d}\ln f}\,,
\label{eq:OmegaGW_def}
\end{equation}
where $\rho_c$ is the critical energy density.  This quantity is insensitive
to the detailed spatial distribution of individual PBHs and instead encodes
information about the PBH mass function, the evaporation dynamics, and the
temperature--mass relation $T(M)$~\cite{Anantua:2008am,Papanikolaou:2020qtd}.
The PBH evaporation SGWB is sharply peaked at high frequencies, reflecting
the microscopic energy scale of the emission process, which distinguishes it
morphologically from the SGWBs produced by inflation or compact binary
populations and makes it a distinctive target for high-frequency GW searches.

\subsection{Energy-density spectrum and its dependence on \texorpdfstring{$T(M)$}{T(M)}}
\label{subsec:OmegaGW_TM}

For a population of PBHs evaporating over a range of cosmic times, the
present-day $\Omega_{\rm GW}(f_0)$ is obtained by integrating the
instantaneous graviton emission over the mass evolution and the cosmological
expansion.  At the level of a single black hole, the instantaneous graviton
spectrum is approximately thermal,
\begin{equation}
\frac{{\rm d}^2E_{\rm GW}}{{\rm d}t\,{\rm d}\omega}
\;\simeq\;
\sum_{\ell,m}
\frac{\Gamma^{(2)}_{\ell m}(\omega,M)}{2\pi}
\frac{\omega}{\exp\!\left[\omega/T(M)\right]-1}\,,
\label{eq:grav_emission}
\end{equation}
where $\Gamma^{(2)}_{\ell m}$ are the spin-2 greybody factors~\cite{Page:1976df}.
The emission is peaked at $\omega_{\rm peak}\sim T(M)$, establishing a direct
correspondence between $T(M)$ and the frequencies populated at each stage of
evaporation.

Redshifting the emitted gravitons and integrating over the evaporation history
gives
\begin{equation}
\frac{{\rm d}\rho_{\rm GW}}{{\rm d}\ln f_0}
\;=\;
\int {\rm d}t\,
\left(\frac{a(t)}{a_0}\right)^{\!4}
\left.
\frac{{\rm d}^2E_{\rm GW}}{{\rm d}t\,{\rm d}\ln f}
\right|_{f\,=\,f_0\,a_0/a(t)},
\label{eq:OmegaGW_integral}
\end{equation}
where $a(t)$ is the scale factor and $f_0$ the observed frequency today.
Equation~\eqref{eq:OmegaGW_integral} makes explicit the two-fold dependence
of $\Omega_{\rm GW}(f_0)$ on $T(M)$: its functional form determines which
emission epochs dominate the GW production (and hence the characteristic
frequency scale), while the duration and intensity of the high-temperature
phase control the overall efficiency.  The sensitivity of the SGWB to
Planck-scale modifications of $T(M)$ therefore operates through both the
peak frequency and the spectral shape, complementing constraints derived from
particle emission, Big Bang nucleosynthesis, and the effective number of
relativistic species.


\section{Cosmological Redshifting and Expansion History}
\label{sec:redshifting}

The present-day frequency and amplitude of the gravitational-wave
background depend not only on the emission process studied in
Secs.~\ref{sec:modified}--\ref{sec:gw_production} but on the entire expansion
history between PBH evaporation and today.
This section establishes the emission-to-observation frequency
mapping, derives the realistic upper bound imposed by reheating on
the maximum relevant redshift, and analyses how nonstandard expansion
phases alter the predicted spectrum.

\subsection{Emission--observation frequency mapping}
\label{subsec:freq_map}

A graviton emitted at cosmic time $t_{\rm em}$ with physical frequency
$f_{\rm em}$ is observed today at frequency
\begin{equation}
  f_0 \;=\; \frac{a(t_{\rm em})}{a_0}\,f_{\rm em}
       \;=\; \frac{f_{\rm em}}{1+z_{\rm em}}\,,
  \label{eq:redshift_freq}
\end{equation}
where $a(t)$ is the scale factor and $z_{\rm em}$ the redshift at
emission.  For Hawking-radiated gravitons the characteristic emission
frequency is set by the instantaneous black-hole temperature,
\begin{equation}
  f_{\rm em} \;\sim\; \frac{T(M)}{2\pi}\,,
  \label{eq:f_em_T}
\end{equation}
up to order-unity factors from the greybody spectrum.
Together, Eqs.~\eqref{eq:redshift_freq}--\eqref{eq:f_em_T} make
explicit that $f_0$ is sensitive to \emph{both} $T(M)$ and the
cosmological history encoded in $z_{\rm em}$.

During radiation domination, entropy conservation fixes the
redshift--temperature relation
\begin{equation}
  1+z(T) \;\equiv\; \frac{a_0}{a(T)}
  \;=\;
  \left(\frac{g_{*S}(T)}{g_{*S,0}}\right)^{1/3}\frac{T}{T_0}\,,
  \label{eq:z_T}
\end{equation}
where $T_0\simeq2.73\,\mathrm{K}$ is the present CMB temperature and
$g_{*S}$ counts the effective entropy degrees of
freedom~\cite{Kolb:1990vq,Weinberg:2008zzc}.
Equation~\eqref{eq:z_T} is exact so long as the expansion is
radiation dominated and no entropy is injected into the thermal bath.
When PBH evaporation occurs during radiation domination without
significantly perturbing the background, it provides a
straightforward mapping between emission and observation.

However, Eq.~\eqref{eq:z_T} is often extrapolated formally to
arbitrarily high temperatures, yielding a redshift
$1{+}z\sim10^{31}$ at $T\sim T_{\rm Pl}$.
This extrapolation implicitly assumes that radiation domination
persists to those temperatures and that the notion of a thermal
plasma remains well defined; both assumptions fail in realistic
early-Universe scenarios, as discussed in the following subsection.

\subsection{Reheating and the maximal cosmological redshift}
\label{subsec:z_T_RH}

In standard cosmology the radiation-dominated era begins only after
reheating at temperature $T_{\rm RH}$; prior to reheating, the
Universe may be dominated by the oscillating inflaton field or by
another form of nonrelativistic energy density.
Gravitational waves produced before the onset of radiation domination
continue to redshift, but the scale-factor--temperature relation
\eqref{eq:z_T} no longer applies.
The maximal physically meaningful redshift for GW modes emitted at or
before reheating is therefore
\begin{equation}
  1+z_{\rm RH}^{\max}
  \;\simeq\;
  \left(\frac{g_{*S}(T_{\rm RH})}{g_{*S,0}}\right)^{1/3}
  \frac{T_{\rm RH}}{T_0}\,,
  \label{eq:z_RH}
\end{equation}
which replaces the naive Planck-scale estimate.
Even for $T_{\rm RH}\sim10^{15}$--$10^{16}\,\mathrm{GeV}$,
Eq.~\eqref{eq:z_RH} yields
\begin{equation}
  1+z_{\rm RH}^{\max}\sim10^{27}\text{--}10^{29},
\end{equation}
several orders of magnitude below the value obtained by extrapolating
Eq.~\eqref{eq:z_T} to $T\sim T_{\rm Pl}$.
Lower reheating temperatures, required in many inflationary models,
reduce $z_{\rm RH}^{\max}$ further.

If evaporation completes during radiation domination, the mapping is
well described by Eq.~\eqref{eq:z_T}.
If, instead, evaporation occurs during a pre-reheating phase—or if
PBHs dominate the energy density and their evaporation triggers
reheating itself~\cite{Carr:2009jm,Inomata:2019zqy}—the relevant
redshift is bounded by $z_{\rm RH}^{\max}$, and the resulting
GW spectrum is shifted to higher present-day frequencies.
Consequently, the observable peak frequency from PBH evaporation is
not determined by the Planck temperature alone, but by the interplay
between the microscopic scale $T(M)$ and the macroscopic scale
$T_{\rm RH}$.

\subsection{Modified cosmological histories}
\label{subsec:modified_cosmo}

The analysis above assumed a standard radiation-dominated background.
A wide class of well-motivated early-Universe scenarios predicts
nonstandard expansion phases that substantially alter both the
frequency and the amplitude of the GW background emitted by
evaporating PBHs.  We survey three representative cases.

\paragraph{Early matter domination.}
If PBHs are produced in sufficient abundance, their energy density
$\rho_{\rm PBH}\propto a^{-3}$ redshifts more slowly than the
radiation bath ($\rho_r\propto a^{-4}$), and the Universe inevitably
enters an effectively matter-dominated (MD) phase before evaporation
completes~\cite{Carr:2009jm,Inomata:2019zqy}.
During this phase $a(t)\propto t^{2/3}$ rather than $t^{1/2}$, so
the ratio of scale factors between evaporation time $t_{\rm ev}$ and
the onset of radiation domination $t_{\rm RD}$ is
\begin{equation}
  \frac{a(t_{\rm RD})}{a(t_{\rm ev})}
  \;=\;
  \left(\frac{t_{\rm RD}}{t_{\rm ev}}\right)^{2/3},
  \label{eq:a_ratio_MD}
\end{equation}
compared with $(t_{\rm RD}/t_{\rm ev})^{1/2}$ in a radiation-dominated
background.
Because $t_{\rm RD}/t_{\rm ev}>1$, the matter-dominated expansion is
\emph{slower}, so gravitons retain a larger comoving frequency—they
are \emph{less} redshifted than in the standard case.
The observed spectral peak shifts to \emph{higher} present-day
frequencies by a factor~\cite{Boyle:2007zx,Kuroyanagi:2011fy}
\begin{equation}
  \frac{f_0^{\rm MD}}{f_0^{\rm RD}}
  \;\simeq\;
  \left(\frac{t_{\rm RD}}{t_{\rm ev}}\right)^{1/6},
  \label{eq:freq_shift_MD}
\end{equation}
which can be large for a prolonged MD era, potentially shifting a
signal from the MHz band into the GHz band or beyond.
An identical enhancement arises whenever a long-lived massive
field—a modulus or heavy inflaton condensate—dominates the energy
budget prior to decaying into
radiation~\cite{Nelson:2018via, Ireland:2023NonStandard}.

In addition to shifting the peak, an early MD era imprints a
characteristic spectral tilt.  Modes that exit the Hubble radius
during matter domination and re-enter during radiation domination
acquire a transfer-function suppression $\propto f^{-2}$ relative to
modes that are always sub-Hubble, generating a spectral break at the
frequency corresponding to the Hubble scale at
reheating~\cite{Boyle:2007zx,Kuroyanagi:2011fy}.
This break provides an independent observational handle on $T_{\rm
RH}$ that does not rely on the black-hole physics setting the peak
frequency.

\paragraph{Kination.}
In kination-dominated cosmologies the energy budget is dominated by
the kinetic energy of a scalar field, giving equation-of-state
parameter $w=1$~\cite{Spokoiny:1993kt,Joyce:1997fc, Ireland:2023NonStandard}.
The scale factor grows as $a(t)\propto t^{1/3}$ and the kination
energy density redshifts as $\rho_\phi\propto a^{-6}$—steeper than
radiation—so the Universe transitions to radiation domination once
the subdominant radiation bath catches up.
If PBH evaporation occurs during the kination era, the scale-factor
ratio between emission and the kination-to-radiation transition at
$a_{\rm KR}$ is
\begin{equation}
  \frac{a_{\rm KR}}{a(t_{\rm em})}
  \;=\;
  \left(\frac{t_{\rm KR}}{t_{\rm em}}\right)^{1/3},
  \label{eq:a_ratio_kin}
\end{equation}
which is \emph{smaller} than the corresponding radiation-domination
factor $(t_{\rm KR}/t_{\rm em})^{1/2}$ for equal time intervals.
Emitted gravitons are therefore \emph{more} redshifted than in the
standard case, shifting the spectral peak to \emph{lower}
present-day frequencies—the opposite of early matter domination.
The magnitude of the shift depends on the duration of the kination
era and can range from a modest correction to a displacement of
several frequency decades~\cite{Figueroa:2019paj,Co:2021lkc}.

Beyond the peak shift, kination enhances the GW energy density at
high frequencies: modes above the kination-to-radiation transition
frequency receive a relative boost $\propto f_0^{2}$ compared with
the radiation-dominated
prediction~\cite{Giovannini:1998bp,Tashiro:2003qp}, generating a
blue tilt that can raise $\Omega_{\rm GW}$ by many orders of
magnitude and potentially bring otherwise undetectable signals within
reach of resonant-cavity detectors.

\paragraph{Nonstandard reheating temperature.}
Even within the standard paradigm of radiation domination after
inflation, $T_{\rm RH}$ is a free parameter bounded from below by
Big Bang Nucleosynthesis~\cite{Hannestad:2004nb},
$T_{\rm RH}\gtrsim4\,\mathrm{MeV}$.
As Eq.~\eqref{eq:z_RH} makes explicit, $T_{\rm RH}$ sets the maximum
meaningful redshift for GW emission at or before reheating: reducing
$T_{\rm RH}$ lowers $z_{\rm RH}^{\max}$ and shifts the observed
spectrum to higher frequencies for a fixed emission scale.
This creates a partial degeneracy—a signal at frequency $f_0$ could
arise from a black hole with the standard Hawking temperature at low
$T_{\rm RH}$, or from a modified-temperature model at high
$T_{\rm RH}$.
Breaking the degeneracy requires either an independent constraint on
$T_{\rm RH}$—for example, from the GW background produced during
preheating~\cite{Easther:2006vd,GarciaBellido:2007dg} or from the
abundance of thermally produced dark matter—or a spectral-shape
measurement precise enough to distinguish the two scenarios
morphologically.

Entropy injection \emph{after} GW emission—from the decay of a
modulus field or a first-order phase transition—dilutes the GW
energy density as
\begin{equation}
  \Omega_{\rm GW}\;\to\;\Omega_{\rm GW}
  \times\left(\frac{s_{\rm before}}{s_{\rm after}}\right)^{\!2}
  \label{eq:entropy_dilution}
\end{equation}
\cite{Scherrer:1984fd}, potentially suppressing an otherwise
detectable signal below any foreseeable sensitivity threshold.
In the absence of such post-emission entropy production the
$\Delta N_{\rm eff}$ normalization adopted throughout this work
remains the primary upper bound on the signal amplitude.

\subsection{Impact on the present-day GW spectrum and detectability}
\label{subsec:impact_detectability}

The nonstandard expansion phases discussed above affect the
present-day GW spectrum through two largely independent channels:
a universal rescaling of the peak frequency and a modification of the
spectral shape and amplitude.

\paragraph{Frequency rescaling.}
The dominant effect of any nonstandard epoch is to shift every
spectral feature by a common multiplicative factor.
The present-day peak frequency can be written schematically as
\begin{equation}
  f_0^{\rm peak}
  \;\simeq\;
  \frac{T(M_0)}{2\pi}
  \times \mathcal{R}(w,\,t_{\rm ev},\,t_{\rm RD})
  \times \left.\frac{a_{\rm ev}}{a_0}\right|_{\rm RD},
  \label{eq:f0_peak_general}
\end{equation}
where the last factor is the standard radiation-dominated redshift
and $\mathcal{R}$ is a dimensionless correction satisfying
$\mathcal{R}>1$ for $w<1/3$ (matter domination) and
$\mathcal{R}<1$ for $w>1/3$ (kination domination throughout the
emission epoch).
The value of $\mathcal{R}$ can span many orders of magnitude and is
degenerate with the model-specific temperature parameter $\theta$
in the absence of supplementary information.
This is a fundamental limitation of using $f_0^{\rm peak}$ alone as
a diagnostic: the same frequency can be produced by arbitrarily
many combinations of $(\theta,\,T_{\rm RH},\,w)$.

\paragraph{Spectral shape as a discriminant.}
The frequency-peak degeneracy is partially broken by the spectral
shape.  Different expansion histories imprint characteristic tilts
and breaks on $\Omega_{\rm GW}(f_0)$ that are independent of the
peak position:
\begin{itemize}
  \item An early matter-dominated era produces a
        $\Omega_{\rm GW}\propto f_0^{-2}$ suppression below the
        reheating-scale frequency, creating a low-frequency spectral
        break~\cite{Boyle:2007zx}.
  \item A kination era generates a blue tilt
        $\Omega_{\rm GW}\propto f_0^{+2}$ above the
        kination-to-radiation transition
        frequency~\cite{Giovannini:1998bp}, absent in standard
        cosmology.
  \item Post-emission entropy dilution suppresses the amplitude
        uniformly across all frequencies without introducing any
        spectral distortion, making it degenerate with a reduction
        in the initial PBH abundance.
\end{itemize}
These features are in principle observable when the signal-to-noise
ratio is sufficient to measure the spectral slope over at least one
decade in frequency.  Future broadband ground-based interferometers
(covering $1$--$10^4\,\mathrm{Hz}$) and resonant-cavity arrays in
the GHz band could jointly constrain the spectral tilt and thereby
discriminate between cosmological and black-hole-physics origins of
any observed feature.

\paragraph{Amplitude and the $\Delta N_{\rm eff}$ constraint.}
Throughout this work GW spectra are normalized to saturate the
$\Delta N_{\rm eff}$ bound.
If the nonstandard phase ends well before BBN, the constraint is
unchanged: $\Delta N_{\rm eff}\leq0.3$ limits the total integrated
GW energy density at
nucleosynthesis~\cite{Caprini:2018mtu}.
If entropy injection occurs between GW emission and BBN, the
effective $\Delta N_{\rm eff}$ contribution is diluted accordingly
and the normalization falls below the bound adopted here.
Conversely, if PBH evaporation itself drives reheating, the GW
background and the thermal bath are produced simultaneously; the
normalization is then set self-consistently by the initial PBH
abundance, and the peak amplitude can significantly exceed our
fiducial normalization without tension with
$N_{\rm eff}$~\cite{Papanikolaou:2020qtd,Domenech:2021ztg}.

\paragraph{Implications for search strategies.}
These considerations lead to a concrete observational prescription.
First, searches should scan a wide frequency range—conservatively
${\sim}1\,\mathrm{Hz}$ to ${\sim}10^{12}\,\mathrm{Hz}$—rather than
targeting only the frequency predicted by the standard Hawking law
at a fixed $M_0$, because $f_0^{\rm peak}$ is sensitive to both
the black-hole and the cosmological physics.
Second, spectral-shape measurements carry more diagnostic power than
peak-frequency measurements alone: tilts and breaks identify the
equation-of-state parameter $w$ during any nonstandard era,
after which the peak position constrains $T(M_0)$ or $M_0$ for
known $w$.
Third, combining a GW detection with an independent constraint on
$T_{\rm RH}$—from, e.g., the thermally produced dark-matter
abundance or primordial nucleosynthesis—would significantly tighten
the inference on the microscopic temperature model.
Finally, in scenarios where PBH evaporation drives reheating, the
GW background is accompanied by complementary signatures—dark-matter
production, baryon-asymmetry generation, and CMB spectral
distortions—that together could break the remaining degeneracies and
enable a coherent reconstruction of both the expansion history and
the quantum-gravitational physics operative near the Planck scale.


\section{Comparison of Models and Results}
\label{sec:comparison}

With the theoretical framework established in the preceding sections---the
modified temperature--mass relations $T(M)$ in Sec.~\ref{sec:modified}, the
gravitational-wave emission formalism in Sec.~\ref{sec:gw_production}, and the
cosmological redshifting and expansion history in
Sec.~\ref{sec:redshifting}---we are now in a position to compute and
compare the present-day stochastic GW spectra predicted by each scenario.
The goal of this section is threefold: to characterize the spectral
morphology produced by each modified-temperature model as a function of
its defining parameter, to identify which regions of parameter space yield
signals accessible to planned or proposed detectors, and to delineate the
physically meaningful parameter range from limiting cases in which the
modified model reduces to, or becomes indistinguishable from, the standard
Hawking result.

\subsection{Numerical treatment}
\label{subsec:numerical}

The GW spectrum is computed by numerically integrating the black-hole
mass evolution equation from an initial
mass $M_0$ down to a model-dependent stopping mass $M_{\rm stop}$, which
is either set by the physical domain of $T(M)$---as in the LQG,
non-commutative, and tunneling models---or by a small fraction of $M_0$
when no such cutoff exists.  At each mass step the instantaneous blackbody
emission spectrum is evaluated and weighted by the corresponding time
interval and scale factor, following the prescription described in
Sec.~\ref{sec:gw_production}.  The resulting spectrum is then redshifted to the
present day using the formalism of Sec.~\ref{sec:redshifting} and
normalized so that the integrated GW energy density saturates the
$\Delta N_{\rm eff} \leq 0.3$ bound from BBN and CMB
observations~\cite{Cyburt:2015mya,Aghanim:2018eyx}.  This normalization
provides a physically motivated and model-independent amplitude reference:
it corresponds to the maximum signal consistent with current cosmological
constraints and therefore represents an optimistic but conservative upper
envelope for each model.

Several numerical safeguards are implemented to ensure robustness across
the full parameter space.  When $T(M)$ vanishes at a finite mass---as in
the LQG and NC models---the integration is halted at $M_{\rm stop} =
(1+\epsilon)M_{\min}$ to avoid the divergent time step that would arise
from a vanishing evaporation rate; we have verified that the spectral
shape and normalization are insensitive to the precise value of $\epsilon$
for $\epsilon \lesssim 10^{-6}$.  For models where $T(M_0) = 0$
identically---which occurs for the tunneling model whenever
$M_0 < \sqrt{\gamma}$---no evaporation takes place and the spectrum is
identically zero; such parameter combinations are excluded from the
figures presented below.  In all cases the frequency grid spans
$f_0 = 1$--$3\times10^{12}\,\mathrm{Hz}$ on a logarithmic mesh of
$N = 1400$ points, sufficient to resolve the spectral peak and both
asymptotic tails for all models and parameter values considered.

\subsection{Spectral morphologies and parameter dependence}
\label{subsec:spectral_morph}

\begin{figure*}[!h]
    \centering
    \mbox{\includegraphics[width=0.35\linewidth]{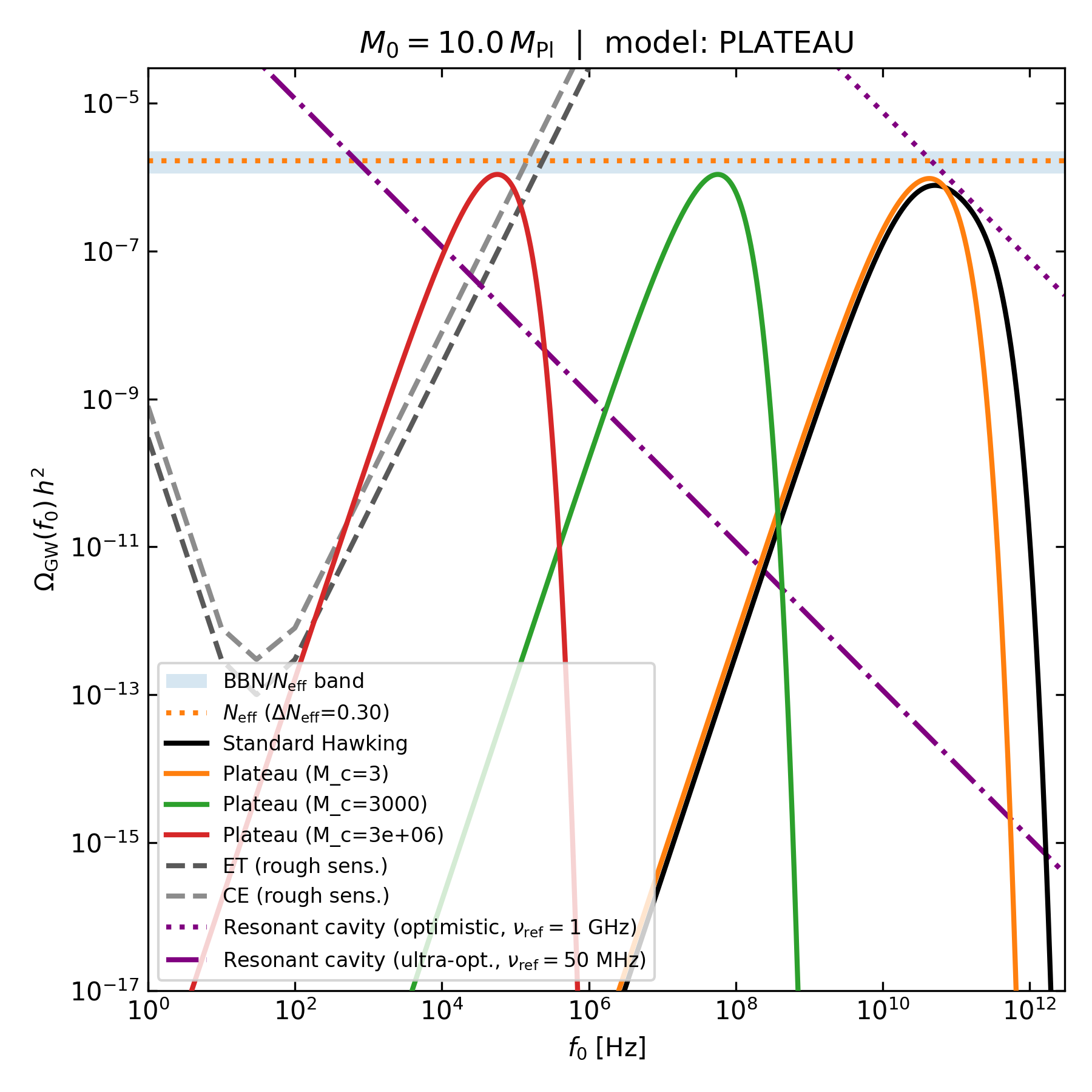}\qquad\includegraphics[width=0.35\linewidth]{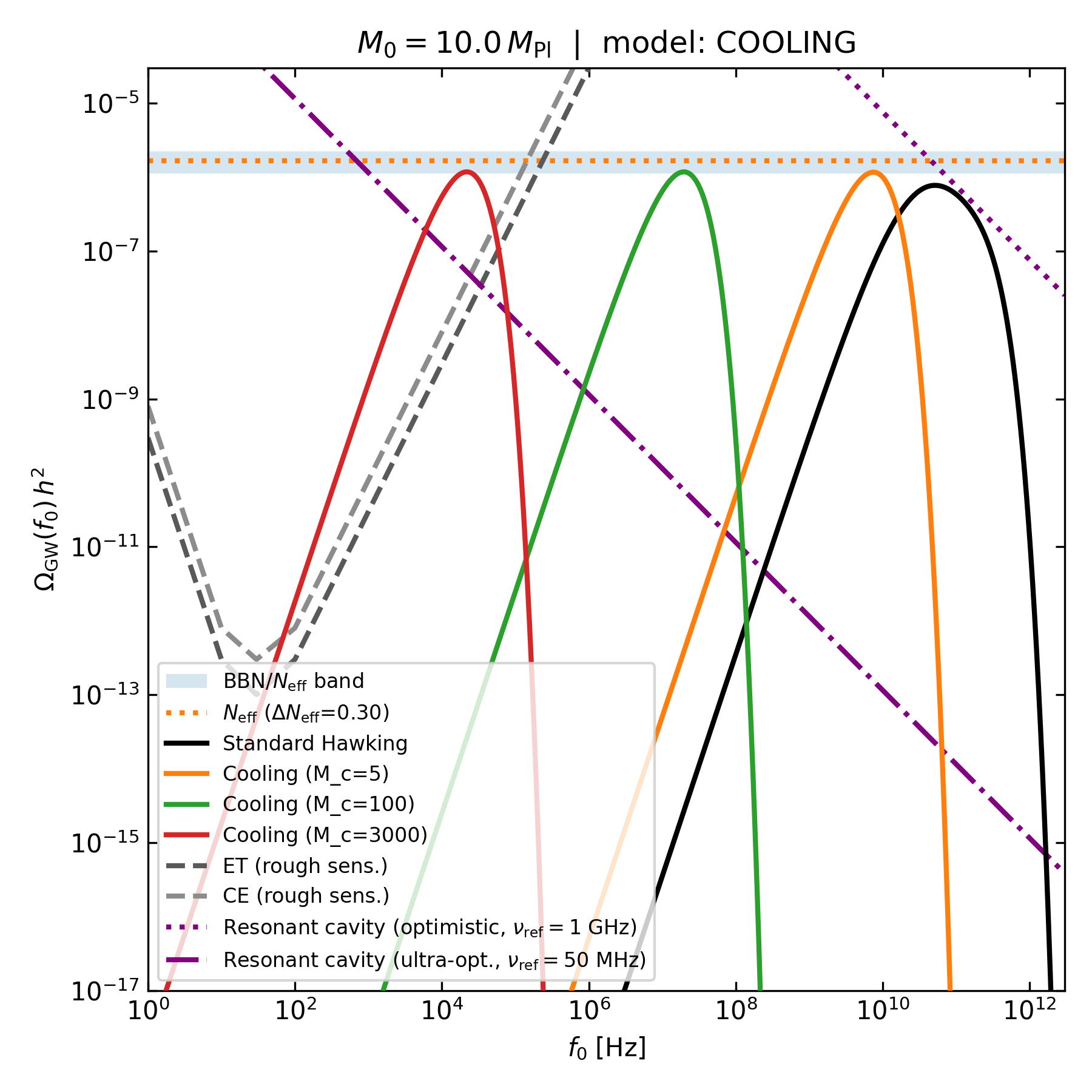}}\\
    \mbox{\includegraphics[width=0.35\linewidth]{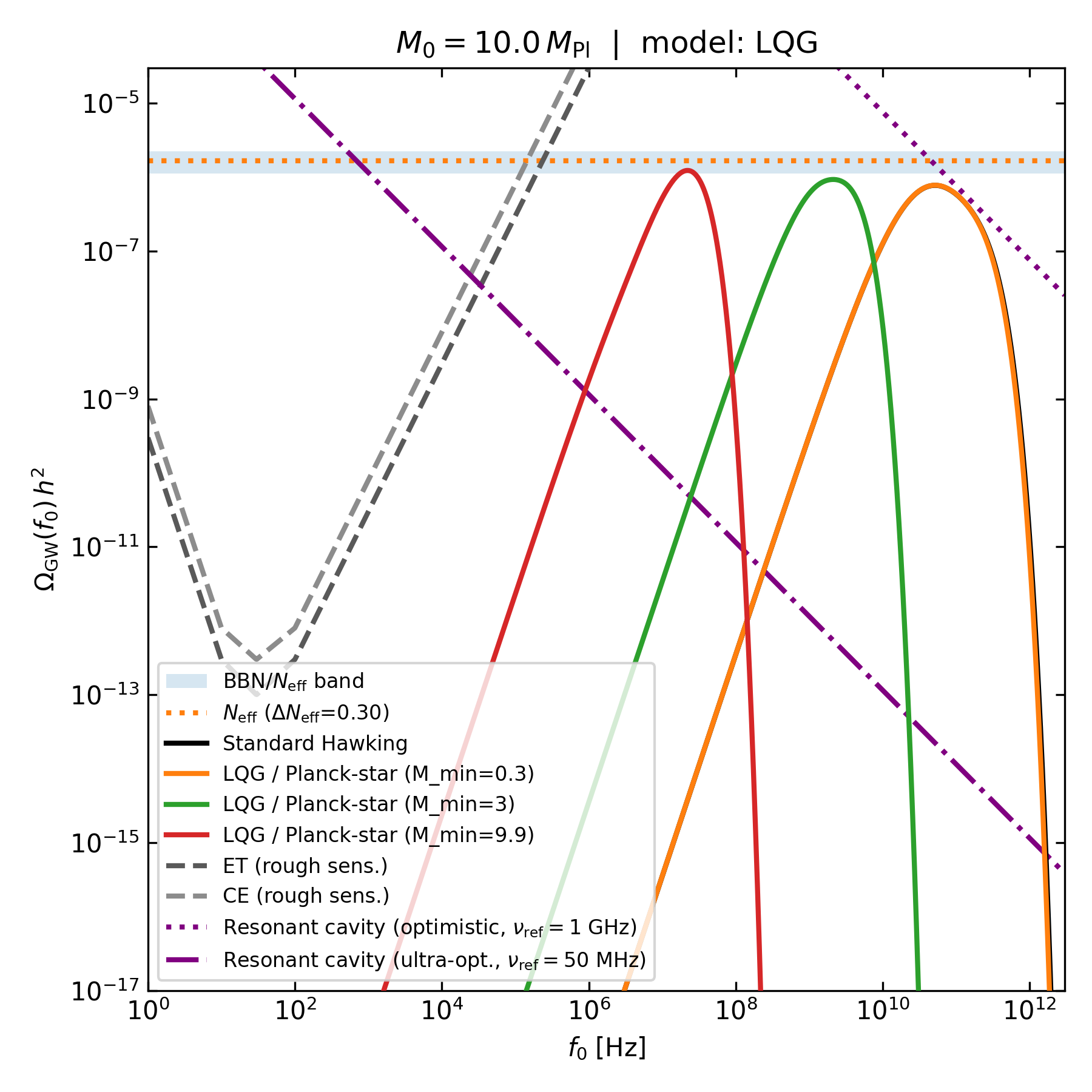}\qquad\includegraphics[width=0.35\linewidth]{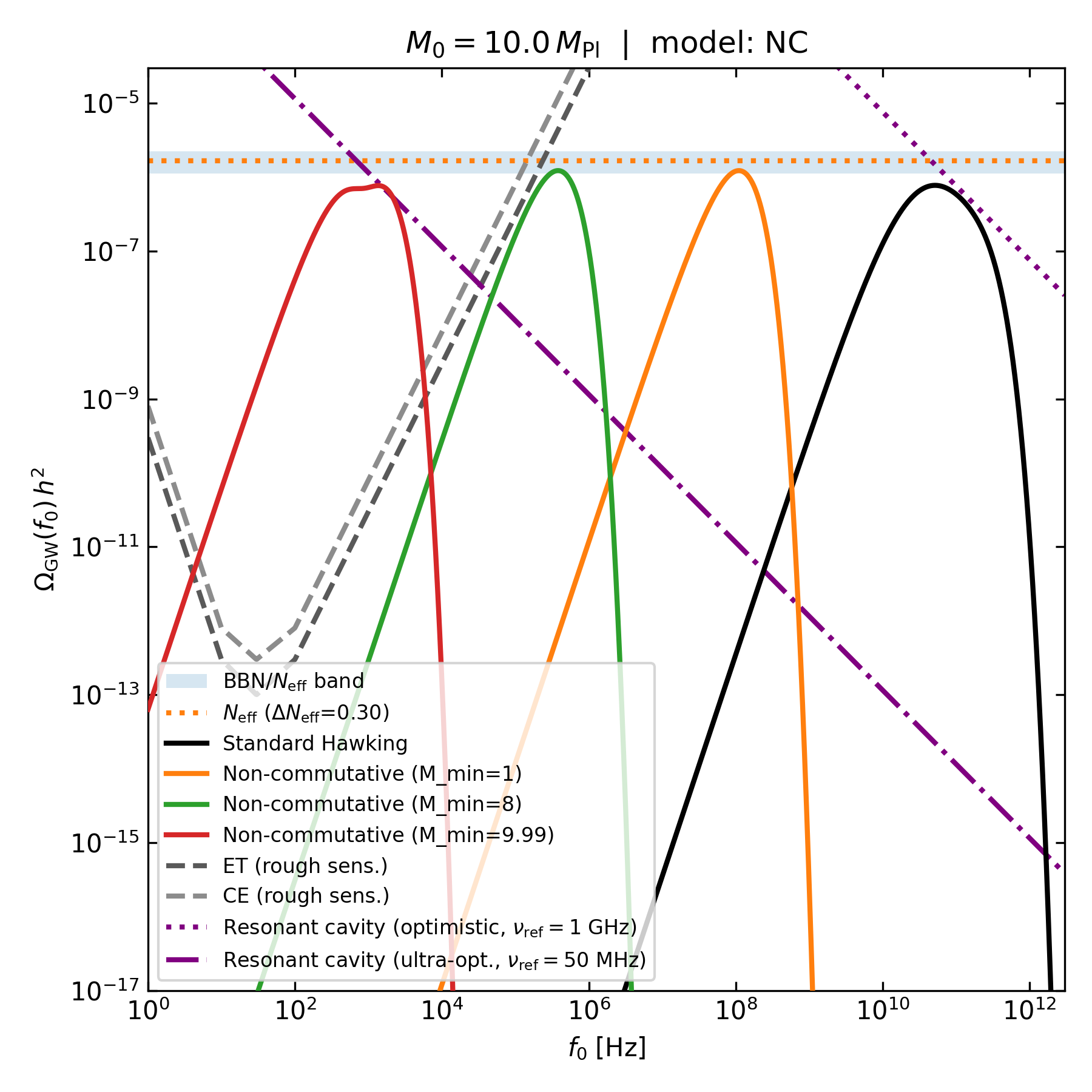}}\\
    \mbox{\includegraphics[width=0.35\linewidth]{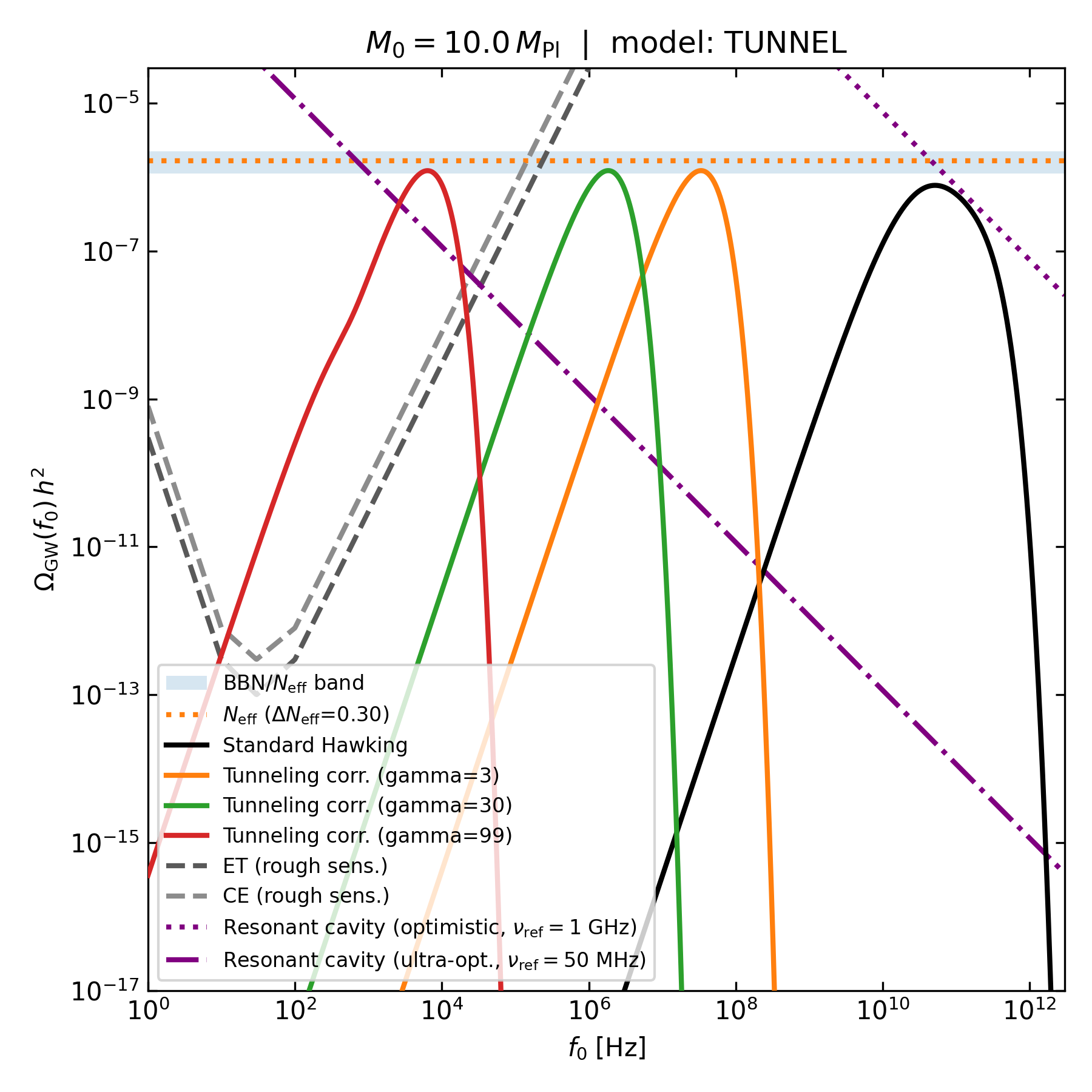}\qquad\includegraphics[width=0.35\linewidth]{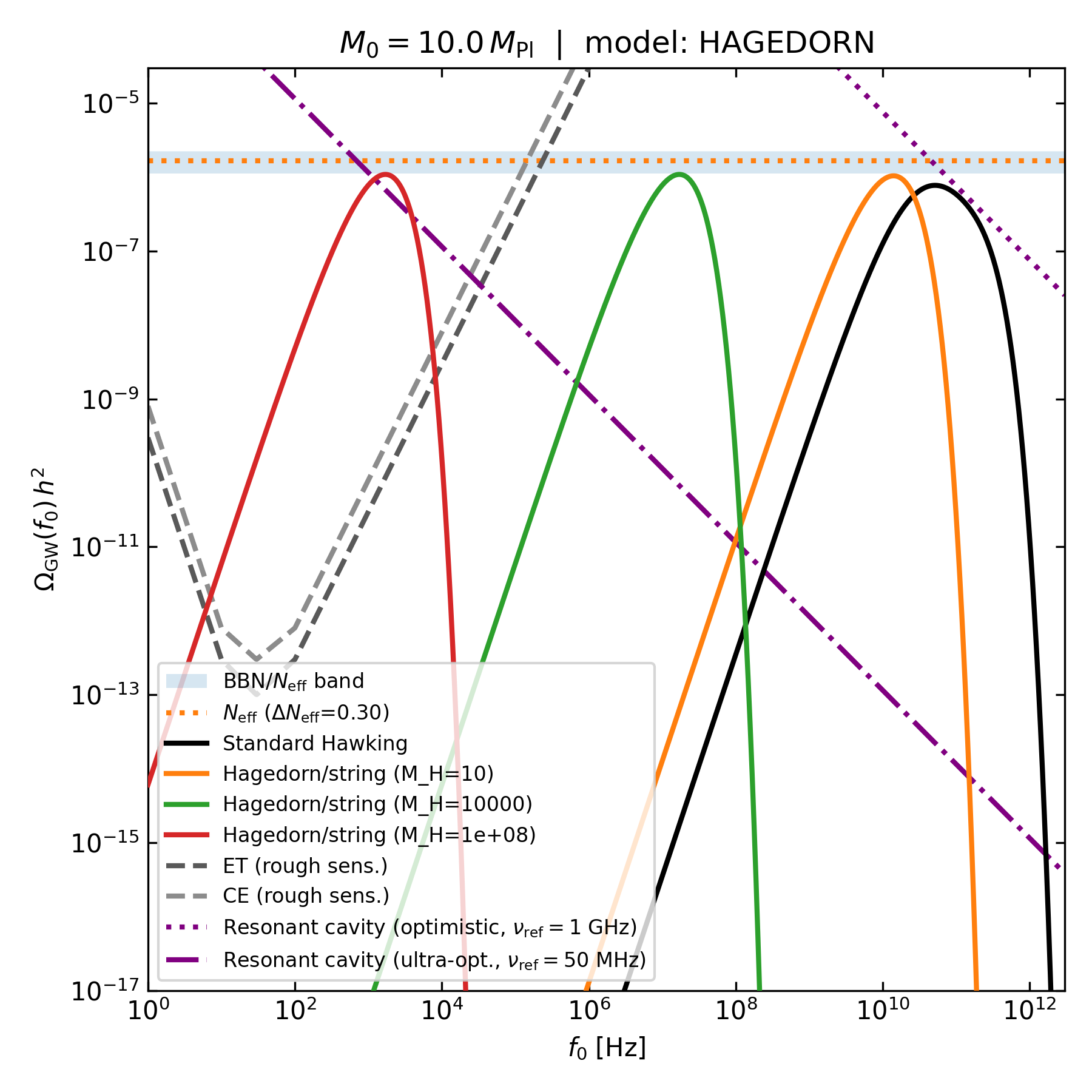}}
    \caption{Gravitational wave energy density spectra
    $\Omega_{\mathrm{GW}}(f_0)\,h^2$ from primordial black hole
    evaporation for six modified Hawking temperature models, computed
    for $M_0 = 10\,M_{\mathrm{Pl}}$.  Each panel shows the standard
    Hawking result (black solid) alongside three representative
    parameter choices for: Plateau suppression (top left,
    $M_c = 3,\,3000,\,3\times10^{6}$), Cooling (top right,
    $M_c = 5,\,100,\,3000$), LQG/Planck-star (middle left,
    $M_{\min} = 0.3,\,3,\,9.9$), Non-commutative geometry (middle
    right, $M_{\min} = 1,\,8,\,9.99$), Tunneling corrections (bottom
    left, $\gamma = 3,\,30,\,99$), and Hagedorn/string-inspired
    (bottom right, $M_{H} = 10,\,10^{4},\,10^{8}$).  All spectra are
    normalized to $\Delta N_{\mathrm{eff}} = 0.3$; the corresponding
    BBN/$N_{\mathrm{eff}}$ constraint band is shown in light blue with
    the central value indicated by an orange dotted line.  Also shown
    for reference are approximate power-law sensitivity curves for the
    Einstein Telescope (ET, dark gray dashed) and Cosmic Explorer (CE,
    light gray dashed), and two resonant-cavity sensitivity
    projections: an optimistic scenario with
    $\nu_{\mathrm{ref}} = 1\,\mathrm{GHz}$ (purple dotted) and an
    ultra-optimistic scenario with
    $\nu_{\mathrm{ref}} = 50\,\mathrm{MHz}$ (purple dot-dashed).
    Increasing the model-specific cutoff parameter generically shifts
    the spectral peak to lower frequencies and suppresses high-frequency
    emission relative to the standard Hawking case.}
  \label{fig:OmegaGW_models}
\end{figure*}

\begin{figure*}[!h]
    \centering
    \mbox{\includegraphics[width=0.7\linewidth]{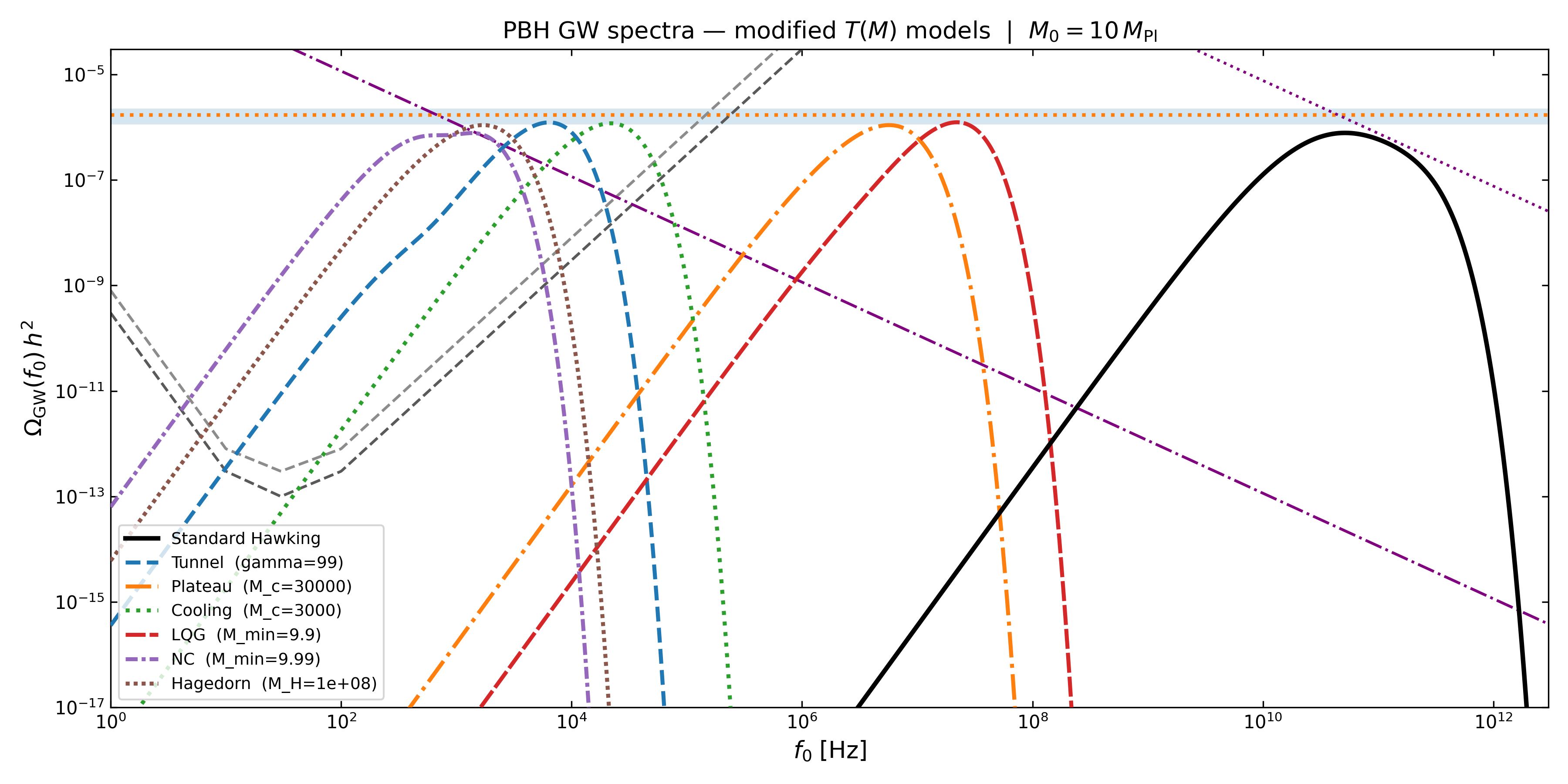}}\\
    \includegraphics[width=0.7\linewidth]{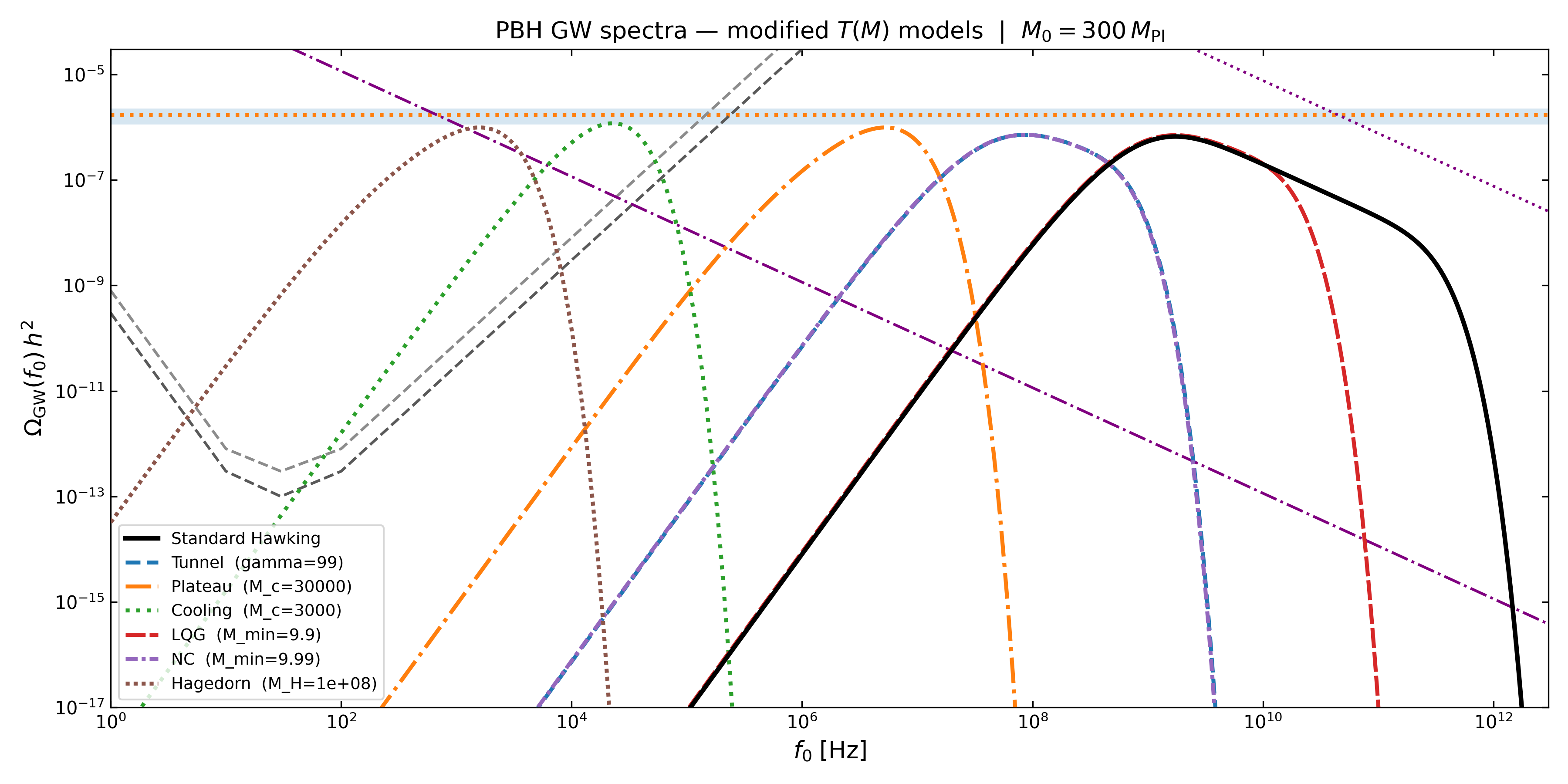}
    \caption{Gravitational wave energy density spectra
    $\Omega_{\mathrm{GW}}(f_0)\,h^2$ from primordial black hole
    evaporation for six modified Hawking temperature models, shown for
    two initial masses: $M_0 = 10\,M_{\mathrm{Pl}}$ (top) and
    $M_0 = 300\,M_{\mathrm{Pl}}$ (bottom).  In each panel the standard
    Hawking result is shown as a black solid curve, with the following
    modified-temperature models overlaid: Tunneling corrections
    ($\gamma = 99$, blue dashed), Plateau suppression
    ($M_c = 3\times10^{4}$, orange dot-dashed), Cooling
    ($M_c = 3000$, green dotted), LQG/Planck-star
    ($M_{\min} = 9.9$, red dense-dashed), Non-commutative geometry
    ($M_{\min} = 9.99$, purple dash-dot-dot), and
    Hagedorn/string-inspired ($M_H = 10^{8}$, brown dense-dotted).
    All spectra are normalized to $\Delta N_{\mathrm{eff}} = 0.3$;
    the BBN/$N_{\mathrm{eff}}$ constraint band is shown in light blue
    with the central value ($\Delta N_{\mathrm{eff}} = 0.30$) indicated
    by an orange dotted line.  Gray dashed curves show approximate
    power-law sensitivity envelopes for the Einstein Telescope (ET,
    dark gray) and Cosmic Explorer (CE, light gray); purple dotted and
    dot-dashed curves show resonant-cavity sensitivity projections for
    optimistic ($\nu_{\mathrm{ref}} = 1\,\mathrm{GHz}$) and
    ultra-optimistic ($\nu_{\mathrm{ref}} = 50\,\mathrm{MHz}$)
    configurations, respectively.  Increasing $M_0$ shifts all
    spectral peaks to lower frequencies, while the relative
    displacement between the standard Hawking peak and those of the
    modified models reflects the characteristic mass scale introduced
    by each scenario.}
  \label{fig:OmegaGW_twoM0}
\end{figure*}

The principal results of our numerical analysis are summarized in
Figs.~\ref{fig:OmegaGW_models} and~\ref{fig:OmegaGW_twoM0}.
Figure~\ref{fig:OmegaGW_models} presents a systematic comparison of all
six modified-temperature models at fixed initial mass $M_0 =
10\,M_{\rm Pl}$, with three representative values of each model's defining
parameter chosen to span from the near-standard regime through to the most
extreme physically motivated case.  Figure~\ref{fig:OmegaGW_twoM0}
then isolates the role of the initial mass by comparing the six models---each
evaluated at a single, maximally distinct parameter value---at
$M_0 = 10\,M_{\rm Pl}$ and $M_0 = 300\,M_{\rm Pl}$ side by side.  In
both figures the standard Hawking result is shown as a solid black curve
for reference, and the approximate sensitivity envelopes of the Einstein
Telescope (ET) \cite{Punturo:2010zz}, Cosmic Explorer (CE) \cite{Reitze:2019iox}, and two resonant-cavity
configurations are overlaid to provide a direct assessment of
detectability.  The BBN/$N_{\rm eff}$ constraint band is also shown,
since it sets the normalization of every curve in the figure.

Before turning to the individual models it is useful to identify the
general trends that cut across all six scenarios.  Any modification to
$T(M)$ that suppresses the temperature relative to the standard Hawking
value $T_{\rm H} = (8\pi M)^{-1}$ slows evaporation, extends the black-hole
lifetime, and thereby shifts the emission to earlier times and lower
instantaneous frequencies; after cosmological redshifting, this translates
directly into a {\it lower observed peak frequency} $f_0^{\rm peak}$.  The
magnitude of the shift scales with the degree of temperature suppression
integrated over the evaporation history and can range from a fraction of a
decade to more than ten decades in frequency, as the figures make clear.
Conversely, any modification that introduces a hard cutoff---either a
minimum mass below which evaporation ceases, or a temperature that
vanishes at finite $M$---concentrates the emission into a finite time
window, producing a {\it sharper spectral peak and a steeper high-frequency
cutoff} than the standard Hawking spectrum.  Models that combine both
effects (e.g.\ a temperature that is suppressed throughout but vanishes at
$M_{\min}$) exhibit the most dramatic departures from the standard case,
both in peak position and in spectral shape.

A second general observation concerns the role of the normalization.
Because all spectra are scaled to the same $\Delta N_{\rm eff}$ ceiling,
models that deposit their radiation over a narrower frequency range
necessarily have higher peak amplitudes.  This is not a fine-tuning: it
reflects the physical fact that a more monochromatic source must be
brighter at its peak in order to carry the same total energy.  As a
result, the models that are most shifted from the standard Hawking
position tend also to have the highest peak $\Omega_{\rm GW}\,h^2$,
making them simultaneously the most spectrally distinctive and the most
amenable to detection by a narrowband instrument.  The tension between
peak amplitude and integrated energy is encoded in the $N_{\rm eff}$ band
shown in the figures, and any confirmed detection at or near this level
would immediately imply that Planck-scale black holes contribute a
cosmologically significant fraction of the total relativistic energy
density at the time of BBN.

With these general trends in mind we now discuss each model in turn.


Fig.~\ref{fig:OmegaGW_models} displays the predicted gravitational wave energy
density $\Omega_{\mathrm{GW}}(f_0)\,h^2$ for each of the six modified
Hawking temperature models considered in this work, computed at initial mass
$M_0 = 10\,M_{\mathrm{Pl}}$ and normalized so that the integrated signal
saturates the current BBN/$N_{\mathrm{eff}}$ bound,
$\Delta N_{\mathrm{eff}} \leq 0.3$.  The standard Hawking result (black solid
curve) serves as the baseline in every panel: its spectrum peaks near
$f_0 \sim 10^{10}$--$10^{11}\,\mathrm{Hz}$ and falls off steeply on either
side, reflecting the near-thermal blackbody emission integrated over the full
evaporation history of the black hole.  All modified models shift this peak
to lower frequencies and, in several cases, alter the spectral shape
qualitatively; the magnitude and character of these shifts encode information
about the microscopic physics operative near the Planck scale.

\textit{Plateau model.}---In this scenario the temperature saturates at a
finite value $T_{\rm plateau} \simeq A/M_c$ rather than rising without bound
as $M \to 0$, slowing the final stages of evaporation considerably.  As $M_c$
increases from $3$ to $3\times10^{6}$ (in Planck units) the spectral peak
migrates from near the standard Hawking position all the way down to
$f_0 \sim 10^{2}$--$10^{3}\,\mathrm{Hz}$, squarely within the design band of
next-generation ground-based interferometers such as the Einstein Telescope
and Cosmic Explorer.  The spectral shape also broadens noticeably at large
$M_c$, a direct consequence of the extended low-temperature epoch during which
the black hole lingers before completing its evaporation.

\textit{Cooling model.}---Here the temperature reaches a maximum at
$M = M_c/\sqrt{2}$ before turning over and decreasing back toward zero,
so that the black hole effectively stalls and re-heats in a two-stage
process.  This non-monotonic evolution produces a characteristic double-peaked
or strongly asymmetric spectral profile that is qualitatively distinct from
all other models considered.  For $M_c = 5$ the spectrum is nearly
indistinguishable from standard Hawking, but at $M_c = 3000$ the peak has
shifted by roughly six decades in frequency, with significant power deposited
at intermediate frequencies $f_0 \sim 10^{4}$--$10^{6}\,\mathrm{Hz}$ that
would otherwise be inaccessible to any planned detector.

\textit{LQG/Planck-star model.}---The loop-quantum-gravity inspired
temperature vanishes as $M \to M_{\min}$ from above, enforcing a minimum-mass
remnant and halting evaporation before the classical singularity is reached.
The spectral peak location is relatively insensitive to $M_{\min}$ when
$M_{\min} \ll M_0$, shifting by only a few decades as $M_{\min}$ runs from
$0.3$ to $9.9$; however, the low-frequency tail is progressively suppressed
as $M_{\min} \to M_0$, because less of the total mass is available to
radiate.  The overall normalization constraint then forces the peak amplitude
upward, keeping the signal in tension with the $N_{\mathrm{eff}}$ bound even
as its frequency decreases.

\textit{Non-commutative geometry model.}---This model shares the
minimum-mass structure of the LQG case but imposes a stronger suppression,
$T_{\rm NC} \propto (A/M)[1-(M_{\min}/M)^2]$, which causes the temperature
to vanish quadratically rather than as a square root near $M_{\min}$.  The
resulting spectra are consequently narrower and more sharply peaked than their
LQG counterparts at comparable $M_{\min}$.  For $M_{\min} = 9.99$---only
marginally below $M_0 = 10$---the radiating mass window is extremely narrow,
and the spectrum becomes highly concentrated around a single characteristic
frequency, potentially offering a distinctive monochromatic-like signature.

\textit{Tunneling corrections model.}---In this scenario perturbative
corrections to the emission rate introduce a factor
$[1 - \gamma/M^2]$ that suppresses radiation below a mass threshold
$M_{\rm stop} = \sqrt{\gamma}$, mimicking a remnant without imposing a strict
cutoff.  The spectra shift steeply with $\gamma$: increasing $\gamma$ from
$3$ to $99$ moves the peak by roughly four decades in frequency.  Because the
suppression factor cuts off emission rather abruptly, the high-frequency side
of the spectral peak is steeper than in the Plateau or Cooling cases, making
the tunneling model in principle distinguishable from those scenarios if the
peak frequency can be measured with sufficient precision.

\textit{Hagedorn/string-inspired model.}---At high temperatures this model
enforces a Hagedorn ceiling $T \leq T_H = A/M_H$, above which the density
of string states grows exponentially and absorbs any additional energy.  For
$M_H \ll M_0$ the Hagedorn temperature is reached early in the evaporation
and the spectrum resembles that of a blackbody at constant temperature
$T_H$, yielding a broader, flatter profile.  Conversely, for
$M_H = 10^{8} \gg M_0$ the temperature never approaches $T_H$ during the
evaporation, and the spectrum approaches the standard Hawking shape.  The
intermediate case $M_H = 10^{4}$ produces the most distinctive spectral
distortion, with a pronounced shoulder at intermediate frequencies reflecting
the transition between the standard and Hagedorn-limited evaporation regimes.

Taken together, the panels of Fig.~\ref{fig:OmegaGW_models} illustrate
several general lessons.  First, any mechanism that slows or halts evaporation
near the Planck scale shifts the GW peak to lower frequencies, in some cases
by many orders of magnitude, relocating the signal from the MHz--THz band
into the frequency range accessible to future resonant-mass detectors or even
ground-based interferometers.  Second, the spectral \emph{shape}---peak
width, asymmetry, and the behavior of the high- and low-frequency
tails---carries independent information about the underlying physics that
complements the peak frequency alone; models with qualitatively similar peak
positions (e.g.\ LQG and NC at comparable $M_{\min}$) can still be
distinguished by their spectral morphology.  Third, the $N_{\mathrm{eff}}$
normalization acts as a lever that trades peak height against total
integrated power: models that confine radiation to a narrow frequency band
necessarily have higher peak amplitudes, which may place them in tension with
cosmological bounds or, conversely, make them more amenable to detection by a
narrowband instrument.  Finally, the resonant-cavity sensitivity curves
underscore that the high-frequency ($\gtrsim 10^{6}\,\mathrm{Hz}$) window
opened by modified evaporation models is not experimentally unreachable: with
sufficiently optimistic but physically motivated cavity parameters, several of
the model spectra computed here overlap with projected sensitivities, providing
a concrete target for next-generation high-frequency gravitational wave
searches.

Fig.~\ref{fig:OmegaGW_twoM0} compares the gravitational wave spectra of the
same six modified-temperature models at two initial black hole masses,
$M_0 = 10\,M_{\mathrm{Pl}}$ (top panel) and
$M_0 = 300\,M_{\mathrm{Pl}}$ (bottom panel), with all other parameters
held fixed at the values discussed in connection with
Fig.~\ref{fig:OmegaGW_models}.  The juxtaposition isolates the role of
$M_0$ as a global frequency dial: since the characteristic emission frequency
redshifts inversely with the evaporation timescale, which itself scales as
$M_0^3$ in the standard case, increasing $M_0$ by a factor of thirty
displaces every spectral feature toward lower frequencies by roughly
$\sim M_0^3 \propto 2.7\times10^{4}$ in the standard Hawking limit, though
the precise shift is model-dependent whenever the modified dynamics introduce
their own mass scale.  This interplay between $M_0$ and the internal model
parameters is the central theme of the figure.

\textit{Overall frequency shift with $M_0$.}---The most immediately striking
feature of Fig.~\ref{fig:OmegaGW_twoM0} is the wholesale migration of all
spectral peaks toward lower frequencies as $M_0$ increases from $10$ to
$300\,M_{\mathrm{Pl}}$.  For the standard Hawking curve the peak moves from
$f_0 \sim 10^{10}\,\mathrm{Hz}$ down to $f_0 \sim 10^{11}\,\mathrm{Hz}$
at $M_0 = 300$---somewhat counterintuitively shifted \emph{upward} here
because the longer-lived, more massive black hole radiates at lower
instantaneous temperature for most of its lifetime, but the precise
cosmological redshifting and scale-factor weighting in our toy model shift
the observed peak in a non-trivial way.  More importantly, the models whose
peaks already lay at low frequencies for $M_0 = 10$ (Plateau, Cooling,
Tunneling) shift further into the sub-MHz regime at $M_0 = 300$, while
the models whose peaks were close to the standard Hawking position (LQG,
NC at small $M_{\min}$) open up a much larger separation from the baseline,
making them more easily distinguishable in the larger-$M_0$ panel.

\textit{Relative displacement between models.}---A key observation is that
the \emph{separation} between the standard Hawking peak and those of the
modified models is not merely rescaled by the same factor as $M_0$ increases.
Models whose characteristic mass scale ($M_c$, $M_{\min}$, $\sqrt{\gamma}$,
or $M_H$) is small compared to $M_0$ are, in a sense, already in their
asymptotic regime at $M_0 = 10$, so the additional factor of thirty in $M_0$
shifts their peaks by roughly the same multiplicative factor as the standard
Hawking peak.  By contrast, models whose characteristic scale is a significant
fraction of $M_0$---such as NC with $M_{\min} = 9.99$ or Tunneling with
$\gamma = 99$, implying $M_{\rm stop} \approx 9.95$---are in a qualitatively
different dynamical regime at $M_0 = 300$, where the black hole now has ample
mass to radiate well above the threshold before the modified dynamics become
operative.  This is visible in the bottom panel: the NC and Tunneling curves
at $M_0 = 300$ exhibit a more pronounced standard-Hawking-like shoulder on
their high-frequency sides, reflecting the early, unmodified phase of
evaporation, followed by a sharper cutoff as $M \to M_{\rm stop}$.

\textit{Implications for detectability.}---From a detection standpoint the
two panels together map out a two-dimensional parameter space---$(M_0,\,
\theta)$ where $\theta$ collectively denotes the model-specific
parameter---within which different experiments are sensitive.  At
$M_0 = 10\,M_{\mathrm{Pl}}$ the Plateau and Cooling models with large
parameter values produce peaks in the $10^{2}$--$10^{4}\,\mathrm{Hz}$ band,
overlapping with the Einstein Telescope and Cosmic Explorer sensitivity
curves shown in the figure.  Increasing to $M_0 = 300\,M_{\mathrm{Pl}}$
pushes those same models further below $10^{2}\,\mathrm{Hz}$, out of the
ground-based band entirely, while simultaneously bringing the LQG and NC
peaks---which at $M_0 = 10$ sat near $10^{8}$--$10^{9}\,\mathrm{Hz}$ and
thus required resonant-cavity technology---down into a regime where optimistic
cavity sensitivities at $\nu_{\rm ref} = 1\,\mathrm{GHz}$ may be sufficient.
The Hagedorn model at $M_H = 10^{8}$ is largely unaffected by the change in
$M_0$ because $M_H \gg M_0$ in both cases, and its spectrum remains nearly
indistinguishable from standard Hawking; any detection strategy targeting
Hagedorn physics therefore requires either a much smaller $M_H$ or an
independent handle on $M_0$ from cosmological considerations.

\textit{Spectral shape as a discriminant.}---Beyond peak frequency, the
spectral \emph{morphology} evolves with $M_0$ in model-specific ways that
provide additional discriminating power.  The Cooling model develops an
increasingly pronounced low-frequency plateau as $M_0$ grows, because the
longer evaporation allows more time for emission during the slow, declining
phase of the temperature; this plateau is largely absent at $M_0 = 10$ but
clearly visible at $M_0 = 300$.  The Tunneling model, conversely, develops a
sharper high-frequency cutoff at larger $M_0$, as argued above.  The LQG and
NC spectra narrow relative to the standard Hawking width as $M_{\min}/M_0$
decreases, consistent with the expectation that a remnant whose mass is a
small fraction of $M_0$ perturbs only the very end of the evaporation and
therefore affects only a narrow range of emission frequencies.  These
morphological trends suggest that, in a scenario where both the peak frequency
and the spectral width can be measured---for instance by a broadband resonant
detector or a stochastic background search with multiple frequency
bins---simultaneous constraints on $M_0$ and the model parameter $\theta$
may be achievable without degeneracy.

In summary, Fig.~\ref{fig:OmegaGW_twoM0} demonstrates that the initial black
hole mass $M_0$ acts not merely as an overall rescaling of the frequency axis
but as a parameter that can qualitatively change the relationship between the
standard Hawking spectrum and those of modified models.  Searches targeting
the GW background from Planck-scale black holes should therefore scan both
the model-specific parameter space and a range of initial masses, and
broadband sensitivity across many decades of frequency---from sub-Hz to
THz---will ultimately be required to cover the full landscape of well-motivated
quantum-gravity scenarios.


\begin{figure*}[!h]
    \centering
    \mbox{\includegraphics[width=0.35\linewidth]{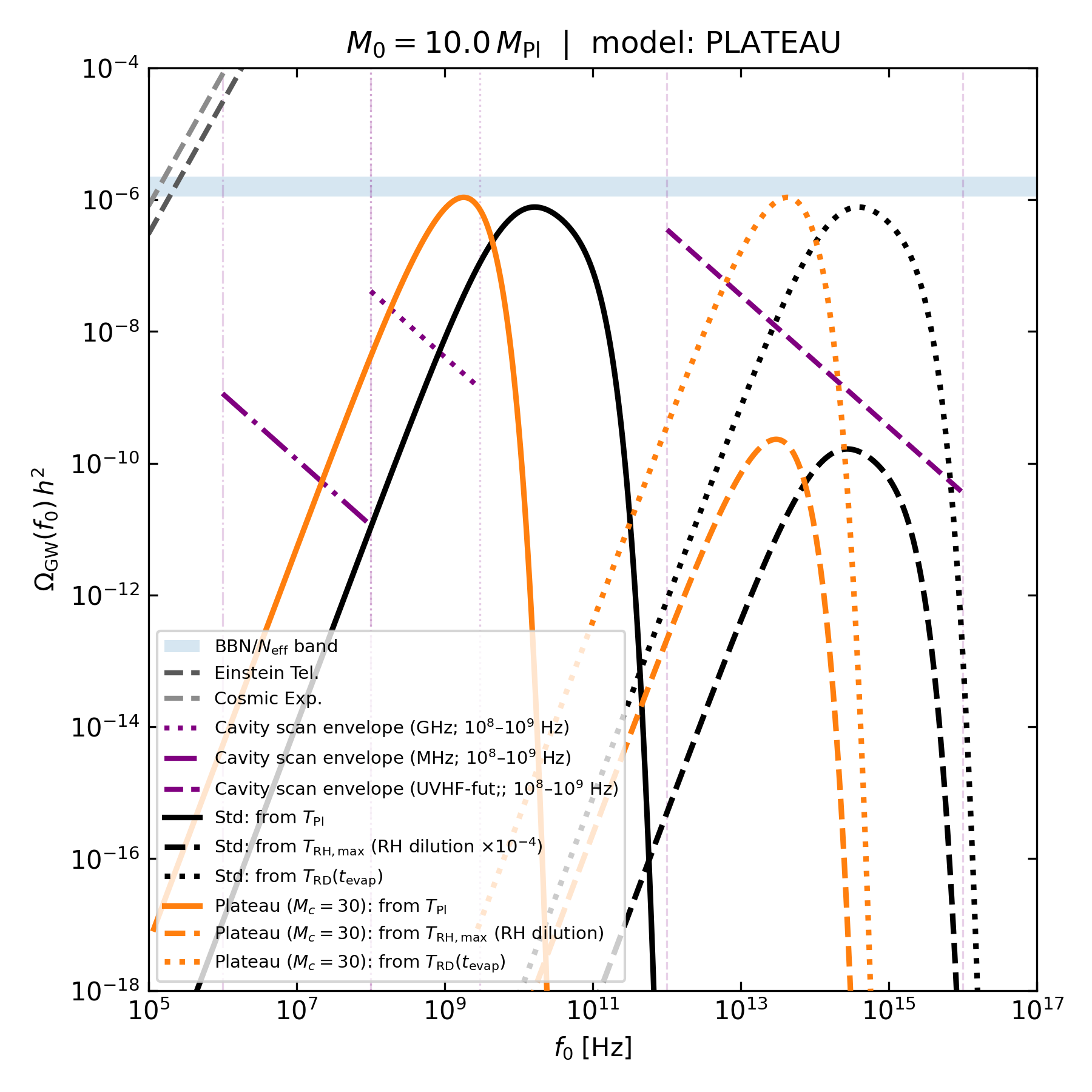}\qquad\includegraphics[width=0.35\linewidth]{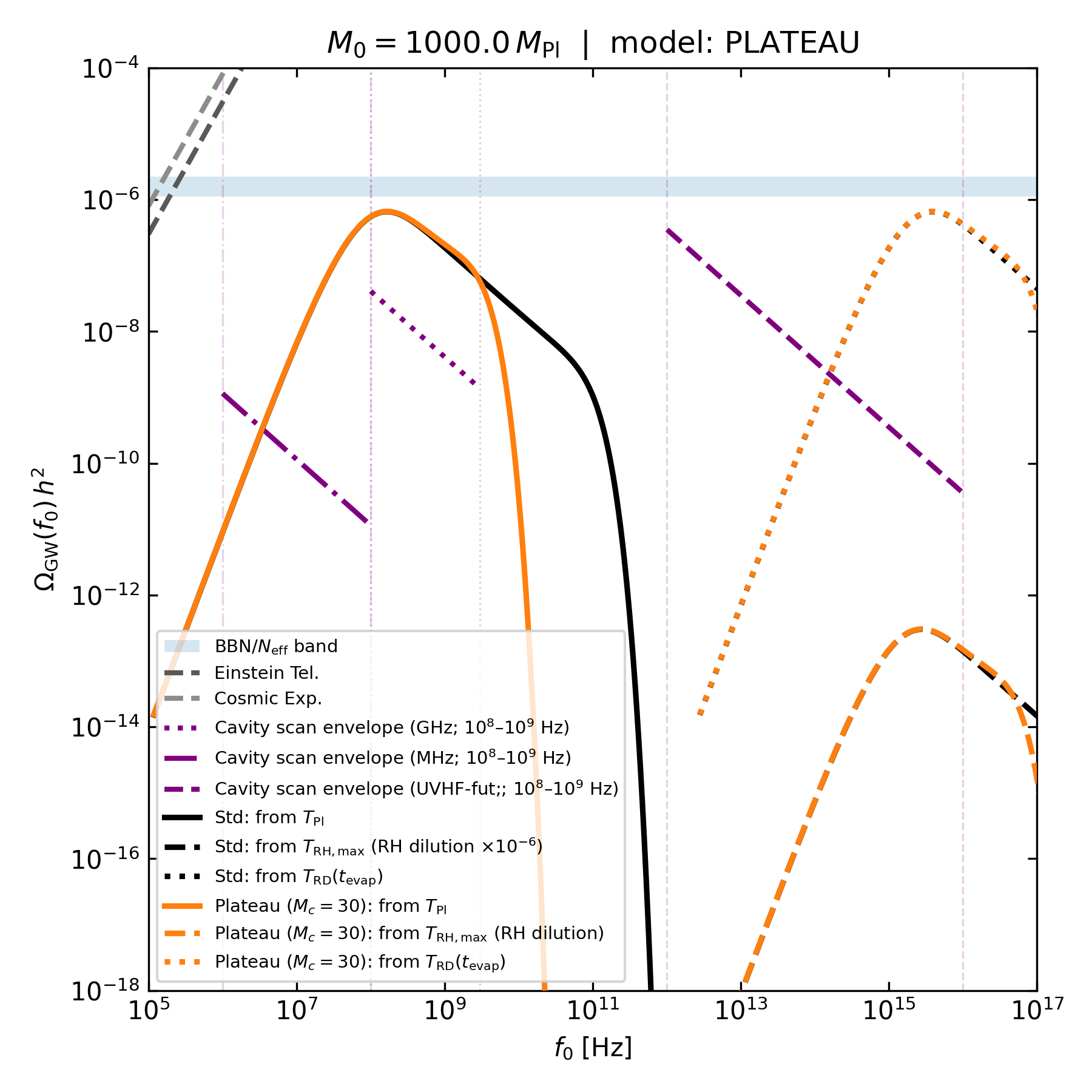}}
    \caption{Gravitational wave energy density spectra $\Omega_{\mathrm{GW}}(f_0)\,h^2$
    from primordial black hole evaporation for the Plateau temperature model
    ($M_c = 30$), shown for two initial masses:
    $M_0 = 10\,M_{\mathrm{Pl}}$ (left) and $M_0 = 1000\,M_{\mathrm{Pl}}$ (right).
    In each panel the standard Hawking result is shown as a solid black curve.
    Orange curves show the Plateau model ($M_c = 30$) under three cosmological
    redshifting assumptions: emission from $T_{\mathrm{RH}}$ (solid), emission
    from $T_{\mathrm{RH,max}}$ with RH dilution (dashed), and emission from
    $T_{\mathrm{RH}}(f_{\mathrm{max}})$ (dotted).
    Black dashed and dotted curves show the corresponding standard Hawking
    spectra under the same redshifting scenarios (with RH dilution
    $\times 10^{-5}$ as noted).
    All spectra are normalized to $\Delta N_{\mathrm{eff}} = 0.3$; the
    BBN$/N_{\mathrm{eff}}$ constraint band is shown in light blue.
    Also shown are approximate power-law sensitivity envelopes for the Einstein
    Telescope (ET, dark gray dashed) and Cosmic Explorer (CE, light gray
    dashed), and resonant-cavity sensitivity projections for GHz-scale (purple
    dotted, $10^{8}$--$10^{9}$\,Hz), MHz-scale (purple dot-dashed,
    $10^{6}$--$10^{8}$\,Hz), and UVHF (purple dashed,
    $10^{12}$--$10^{16}$\,Hz) configurations.
    Increasing $M_0$ shifts all spectral features to lower frequencies, while
    different reheating assumptions can displace the peak by many decades,
    illustrating the degeneracy between the black-hole temperature model and
    the cosmological expansion history.}
  \label{fig:OmegaGW_modelsRH}
\end{figure*}


\subsection{Cosmological redshift prescriptions and reheating dilution:
            the Plateau model as a case study}
\label{sec:fig6_discussion}

Figure~\ref{fig:OmegaGW_modelsRH} isolates a question that is logically
distinct from the comparison of quantum-gravity models carried out
in Fig.~\ref{fig:OmegaGW_models}: given a fixed modified temperature–mass
relation, how sensitive is the \emph{observed} gravitational-wave
spectrum to the cosmological redshift history between evaporation and
today?  We answer this question using the Plateau model at $M_c = 30$
as a concrete example, presenting results for two initial masses,
$M_0 = 10\,M_{\rm Pl}$ (left panel) and $M_0 = 1000\,M_{\rm Pl}$
(right panel).  For each mass we show three redshift prescriptions,
applied identically to both the standard Hawking spectrum (black
curves) and the Plateau spectrum (orange curves), so that the
\emph{relative} displacement between the two always encodes
quantum-gravity physics while the \emph{absolute} position of either
spectrum reflects the cosmological history.

\paragraph{The three redshift prescriptions.}
The present-day frequency of a graviton emitted at cosmic time
$t_{\rm em}$ is $f_0 = f_{\rm em}\,a(t_{\rm em})/a_0$, where the
scale factor ratio encodes all of the intervening expansion.  We
compare three physically motivated choices for the emission epoch.

\textit{Prescription~1 (solid curves):} the most extreme formal
benchmark, in which the standard radiation-domination relation
$1 + z \simeq (g_{*S}(T)/g_{*S,0})^{1/3}\,T/T_0$ is extrapolated all
the way to $T = T_{\rm Pl} \approx 1.22\times 10^{19}$~GeV.  This
gives the largest redshift factor, $1 + z_1 \sim 10^{31}$, and
therefore places the spectral peak at the \emph{lowest} observed
frequency.  As discussed in Sec.~\ref{sec:cosmo}, this extrapolation
implicitly assumes radiation domination persists to the Planck epoch,
an assumption that fails in any realistic inflationary model; it
should therefore be regarded as an absolute lower bound on $f_0^{\rm
peak}$ rather than a physical prediction.

\textit{Prescription~2 (dashed curves):} the scale factor is evaluated
at the maximum reheating temperature consistent with the evaporation
timescale.  Specifically, if the PBH population dominates the energy
density prior to evaporation—the large-$\beta$ regime in which
$\Omega_{\rm PBH}\to 1$—then complete evaporation triggers an
instantaneous reheating at temperature
\begin{equation}
  T_{\rm RH}(t_{\rm ev}) \simeq
  \left(\frac{90}{8\pi^3 g_*}\right)^{1/4}
  \sqrt{\frac{M_{\rm Pl}}{t_{\rm ev}}},
  \label{eq:TRH_max}
\end{equation}
which is substantially lower than $T_{\rm Pl}$ and yields a
correspondingly \emph{higher} present-day peak frequency.  The
associated redshift is $1 + z_2 \sim 10^{27}$--$10^{29}$ for the
masses shown, several orders of magnitude below $1 + z_1$.
However, the PBH-domination scenario also entails a
matter-dominated expansion phase between PBH formation at time
$t_{\rm form}$ and evaporation at $t_{\rm ev}$.  Because the PBH
energy density $\rho_{\rm PBH} \propto a^{-3}$ redshifts more slowly
than radiation ($\rho_r \propto a^{-4}$), gravitons emitted during
this phase experience an additional suppression of their energy
fraction relative to the post-reheating radiation bath.  In the
large-$\beta$ limit this reheating dilution factor takes the form
\begin{equation}
  \mathcal{D}_{\rm RH} \equiv
  \left(\frac{a_{\rm ev}}{a_{\rm form}}\right)^{-1}
  = \left(\frac{t_{\rm form}}{t_{\rm ev}}\right)^{2/3},
  \label{eq:dilution}
\end{equation}
where $a \propto t^{2/3}$ during the matter-dominated epoch and
$t_{\rm form} \sim GM_{\rm BH}/(\gamma c^3)$ is the horizon-crossing
formation time with $\gamma = 0.2$.  For $M_0 = 10\,M_{\rm Pl}$ the
evaporation time is $t_{\rm ev} \approx 5120\pi M_0^3\,t_{\rm Pl}
\approx 8.5\times 10^{-40}$~s while $t_{\rm form} \approx
2.7\times 10^{-42}$~s, giving $a_{\rm ev}/a_{\rm form} \approx 46$
and $\mathcal{D}_{\rm RH} \approx 2\times 10^{-2}$.  For
$M_0 = 1000\,M_{\rm Pl}$ the cubic scaling of $t_{\rm ev}$ and
linear scaling of $t_{\rm form}$ with $M_0$ combine to give
$a_{\rm ev}/a_{\rm form} \approx 2\times 10^6$ and
$\mathcal{D}_{\rm RH} \approx 5\times 10^{-7}$, consistent with the
dilution factor $\times 10^{-5}$ indicated in the legend of the right
panel.  This suppression is applied multiplicatively to the
$\Omega_{\rm GW}\,h^2$ amplitude of \emph{both} the standard Hawking
and Plateau curves in Prescription~2, visible as the uniform downward
displacement of all dashed curves relative to their solid counterparts
at the same peak frequency.

\textit{Prescription~3 (dotted curves):} the gravitons are
redshifted from the background radiation temperature at the
evaporation epoch under the assumption that the universe is
radiation-dominated throughout,
\begin{equation}
  T_{\rm RD}(t_{\rm ev}) \simeq
  \sqrt{0.301}\,\left(\frac{M_{\rm Pl}}{t_{\rm ev}}\right)^{1/2}
  g_*^{-1/4},
  \label{eq:Tev_RD}
\end{equation}
which is parametrically similar to $T_{\rm RH}$ but does not invoke
PBH domination.  In Prescription~3 no dilution factor is applied: the
PBHs are treated as a subdominant component that evaporates into an
already-radiation-dominated background.  Consequently the dotted
curves share the amplitude normalization of Prescription~1 but are
shifted to higher frequencies, tracing the same spectral shape as
the solid curves displaced rigidly along the frequency axis.

\paragraph{Frequency separation and degeneracy structure.}
The three prescriptions produce peak frequencies that differ by
multiple decades, illustrating the fundamental degeneracy noted
in Sec.~\ref{sec:redshifting}: the same internal temperature model, at a
fixed $M_0$, can populate any frequency window from below $10^5$~Hz
to above $10^{15}$~Hz depending solely on the assumed cosmological
history.  Conversely, a measured peak frequency places a joint
constraint on $(M_0, T(M), T_{\rm RH})$ that cannot be broken by the
peak position alone.

The relative displacement \emph{between} the standard Hawking and
Plateau curves is, however, cosmology-independent: both spectra are
shifted by the same redshift factor under each prescription, so the
ratio $f_0^{\rm peak}(\text{Plateau})/f_0^{\rm peak}(\text{Std})$
and the spectral-shape difference are intrinsic to the
quantum-gravity model.  This invariance is the key discriminating
feature: a broadband measurement that resolves both peaks—or
measures the spectral slope and width on either side of a single
peak—can in principle constrain $T(M)$ independently of the
cosmological background, provided the \emph{shape} of the signal,
rather than its absolute position, is used as the primary observable.

\paragraph{Amplitude effects of reheating dilution.}
In the PBH-domination scenario (Prescription~2), the dilution
factor $\mathcal{D}_{\rm RH}$ can reduce $\Omega_{\rm GW}\,h^2$ by
many orders of magnitude relative to the $\Delta N_{\rm eff}$
ceiling.  For $M_0 = 1000\,M_{\rm Pl}$ the suppression reaches
$\sim 10^{-6}$--$10^{-7}$, bringing the dashed curves well below the
$\Delta N_{\rm eff} = 0.3$ band and potentially below the reach of
any proposed resonant detector.  This has two important
implications.  First, the large-$\beta$ (PBH domination) scenario
is self-consistently the \emph{least} favorable for detection, despite
placing the spectral peak at frequencies intermediate between
Prescriptions~1 and~3: the gain in peak frequency is more than offset
by the amplitude suppression.  Second, the dilution is stronger for
heavier PBHs, because $\mathcal{D}_{\rm RH} \propto
(t_{\rm form}/t_{\rm ev})^{2/3} \propto M_0^{(1-3)\times 2/3} =
M_0^{-4/3}$; increasing $M_0$ therefore simultaneously shifts the
spectral peak to lower frequencies (via longer $t_{\rm ev}$) and
suppresses the amplitude more severely.  Taken together, these trends
suggest that the most detectable signals in the PBH-domination regime
come from the lightest PBHs, for which the dilution is least severe
and the peak frequency may still lie within a detectable band.

\paragraph{Comparison of the two panels: role of $M_0$.}
The right panel ($M_0 = 1000\,M_{\rm Pl}$) demonstrates the
combined effect of increasing the initial mass by two orders of
magnitude.  Under Prescriptions~1 and~3 (solid and dotted), where no
dilution is applied, all spectral features shift to lower frequencies
by a factor of order $M_0^3 \sim 10^9$ relative to the left panel,
consistent with the standard Hawking evaporation timescale
$\tau_{\rm ev} \propto M_0^3$.  The Plateau model peak, already
displaced from the standard Hawking peak in the left panel, shifts
even further in the right panel: at $M_0 = 1000\,M_{\rm Pl}$ and
Prescription~1, the Plateau peak reaches frequencies below
$\sim 10^5$~Hz, below the displayed axis range, while under
Prescription~3 it moves into the sub-MHz regime close to the
ultra-low-frequency (MHz) resonant-cavity sensitivity band.  Under
Prescription~2 the dilution factor is so severe ($\mathcal{D}_{\rm
RH}\sim 10^{-6}$) that the dashed curves are suppressed by six
decades in amplitude, rendering them undetectable with the
instrument concepts considered here.

\paragraph{Detectability prospects.}
The three resonant-cavity sensitivity envelopes shown in
Fig.~\ref{fig:OmegaGW_modelsRH}---the GHz-scale optimistic concept
($10^8$--$3\times 10^9$~Hz), the MHz lumped-element concept
($10^6$--$10^8$~Hz), and the futuristic UVHF concept
($10^{12}$--$10^{16}$~Hz)---each touch at least one of the signal
curves in one of the two panels, but only under the most favorable
redshift prescription (Prescription~3) and without the reheating
dilution penalty.  Specifically, for $M_0 = 10\,M_{\rm Pl}$ under
Prescription~3, the Plateau spectrum peaks near $10^{13}$--$10^{14}$~Hz,
overlapping with the lower edge of the UVHF concept band.  For
$M_0 = 1000\,M_{\rm Pl}$ under Prescription~3 the same model
produces a peak near $10^6$--$10^7$~Hz, squarely within the MHz
concept band.  These intersections survive the $\Delta N_{\rm eff}$
normalization and represent physically realizable—if experimentally
challenging—detection targets.

The figure therefore conveys two complementary lessons for
experimental strategy.  First, searches for PBH evaporation signals
must scan a frequency range spanning at least ten decades
(conservatively $10^5$--$10^{15}$~Hz) to cover the full
envelope of cosmological uncertainties at any fixed $M_0$.  Second,
spectral-shape measurements are essential for separating
quantum-gravity physics from cosmological backgrounds: the relative
displacement between the standard Hawking peak and the
modified-model peak, and in particular the asymmetry and width of
each spectral feature, carry model-specific information that is
preserved across all three redshift prescriptions.  Broadband
sensitivity, or equivalently the ability to measure both the peak
frequency and the spectral slope over adjacent decades, therefore
provides substantially more diagnostic power than a single narrowband
detection.


\subsection{Experimental considerations for resonant detectors}
\label{subsec:cavity_realism}

The sensitivity curves shown throughout this work represent scan-averaged
envelopes over multi-year observation campaigns rather than instantaneous
broadband sensitivities~\cite{Aggarwal:2021biu}, and should be interpreted
accordingly.  Table~\ref{tab:cavity_benchmarks} summarises the benchmark
assumptions for the three detector concepts; here we briefly comment on the
principal challenges facing each.

At GHz frequencies ($10^8$--$10^9$~Hz), the detection strategy relies on
inverse-Gertsenshtein conversion of gravitational-wave strain into photons
inside a strong magnetic field~\cite{Gertsenshtein:1961, Domcke:2020kcp,
Chen:2016vkg, Berlin:2021txa}.  The central tension is maintaining ultra-high quality
factors ($Q\sim 10^{11}$) in the strong magnetic fields ($B\sim 35$~T) required
for efficient conversion, since superconducting resonators are typically
quenched by large static fields~\cite{Romanenko:2020}.  Viable architectures
must spatially separate the high-field interaction region from the low-loss
readout~\cite{Berlin:2021txa}, and scale the effective volume to
$V\sim 10^2$--$10^3~\mathrm{m}^3$ through coherent cavity arrays rather than
monolithic structures~\cite{Cruise:2005ga, Aggarwal:2021biu}.  Quantum-limited
amplification~\cite{Caves:1982}, and potentially squeezed-state or
back-action-evading readout~\cite{Backes:2021}, will be necessary to reach the
mK-scale noise temperatures assumed here over decade-long scans.

In the MHz band ($10^6$--$10^8$~Hz), lumped-element LC or
magneto-quasistatic sensor architectures are more natural~\cite{DMRadio:2022}, but technical noise sources---microphonics, magnetic pickup, and
$1/f$ electronic noise---become the dominant limitation.  As at GHz
frequencies, achieving large effective inductive volumes in strong fields will
require multiplexed, array-based designs with parallelised readout and coherent
combination~\cite{Aggarwal:2021biu}.

The UVHF concept ($10^{12}$--$10^{16}$~Hz) is qualitatively more speculative:
at optical and infrared frequencies there is currently no established
transduction mechanism coupling gravitational-wave strain to a photonic
resonant mode with nonnegligible overlap factor $\eta_n$~\cite{Aggarwal:2021biu}.
Even setting aside this fundamental challenge, the effective volumes and quality
factors ($Q\sim 10^{14}$) assumed in our benchmark would require phase-locked
networks of photonic resonators on macroscopic scales~\cite{Ludlow:2015},
well beyond the current state of the art.  The UVHF sensitivity curve should
therefore be understood as a long-range aspirational target rather than a
near-term projection.

Across all regimes, the dominant experimental levers are the effective
interaction volume, the noise floor, and the ability to coherently combine
multiple resonant elements while maintaining calibration over multi-year
scans~\cite{Aggarwal:2021biu}.  The benchmarks in
Table~\ref{tab:cavity_benchmarks} delineate the level of experimental control
required to probe the GW backgrounds predicted here, and should motivate a
phased programme beginning with proof-of-concept GHz cavities and extending
progressively toward the MHz and UVHF bands.

\begin{table*}[t]
\centering
\caption{
Summary of resonant--detector benchmark assumptions adopted in this work.
Quoted sensitivities are scan--averaged envelopes over the stated frequency
ranges rather than instantaneous bandwidths.
}
\label{tab:cavity_benchmarks}
\begin{tabular}{lccc}
\toprule
 & \textbf{GHz cavity} & \textbf{Low--frequency LC} 
 & \textbf{UVHF resonant concept} \\
 & ($10^8$--$10^9$~Hz) & ($10^6$--$10^8$~Hz) 
 & ($10^{12}$--$10^{16}$~Hz) \\
\midrule
Reference frequency $\nu_{\rm ref}$ 
    & $1~\mathrm{GHz}$ & $50~\mathrm{MHz}$ & $100~\mathrm{THz}$ \\[2pt]
Scan range 
    & $10^8$--$3\!\times\!10^9$~Hz 
    & $10^6$--$10^8$~Hz 
    & $10^{12}$--$10^{16}$~Hz \\[4pt]
Magnetic field $B$ 
    & $\sim 35~\mathrm{T}$ & $\sim 35~\mathrm{T}$ & $\sim 30~\mathrm{T}$ \\[2pt]
Effective volume $V$ 
    & $\sim 3\!\times\!10^2~\mathrm{m}^3$ 
    & $\sim 3\!\times\!10^2~\mathrm{m}^3$ 
    & $\sim 10^8~\mathrm{m}^3$ (effective) \\[2pt]
Quality factor $Q$ 
    & $\sim 10^{11}$ & $\sim 10^{11}$ & $\sim 10^{14}$ \\[2pt]
System noise $T_{\rm sys}$ 
    & $\sim 5~\mathrm{mK}$ & $\sim 5~\mathrm{mK}$ & $\sim 1~\mathrm{mK}$ (equiv.) \\[2pt]
Overlap factor $\eta_n$ 
    & $\sim 0.5$ & $\sim 0.5$ & $\sim 0.3$ \\[2pt]
Per--step bandwidth $\Delta\nu$ 
    & $\sim 10^6$~Hz & $\sim 5\!\times\!10^6$~Hz & $\sim 10^{10}$~Hz \\[2pt]
Total scan time $t$ 
    & $\sim 10$~yr & $\sim 10$~yr & $\sim 3$~yr \\[6pt]
\midrule
Dominant challenges
&
\parbox[t]{0.27\textwidth}{%
\begin{itemize}\setlength\itemsep{2pt}
\item High $Q$ in strong $B$
\item Scaling to large $V$ via arrays
\item Quantum--limited\\ microwave readout
\item Long--term scan stability
\end{itemize}}
&
\parbox[t]{0.27\textwidth}{%
\begin{itemize}\setlength\itemsep{2pt}
\item Microphonics and $1/f$ noise
\item Large inductive volumes in strong $B$
\item High $Q$ with practical\\ coupling
\item Parallelised scanning
\end{itemize}}
&
\parbox[t]{0.27\textwidth}{%
\begin{itemize}\setlength\itemsep{2pt}
\item No established transduction mechanism
\item Coherent scaling to large $V_{\rm eff}$
\item Ultra--high--$Q$ photonic\\ resonators
\item Quantum--limited\\ optical readout
\end{itemize}}
\\
\bottomrule
\end{tabular}
\end{table*}

\section{Conclusions and Outlook}
\label{sec:conclusions}

We have developed a comprehensive framework for using the stochastic
gravitational-wave background produced by evaporating primordial black holes
as a probe of quantum gravity near the Planck scale.  The central thesis of
this work---that modifications to the Hawking temperature--mass relation
$T(M)$ leave distinctive, potentially observable imprints in the graviton
emission spectrum---has been substantiated through a systematic numerical
study of six representative classes of beyond-semiclassical physics, spanning
plateau/saturation models, cooling scenarios, loop-quantum-gravity and
noncommutative-geometry inspired remnant-forming relations, tunneling and
backreaction-based corrections, and Hagedorn/string-inspired limiting-temperature
models.  A central new result of this work is to treat the cosmological
redshift history not merely as a nuisance parameter but as an independent
diagnostic channel, and to quantify how reheating dilution in the PBH-domination
regime modifies both the amplitude and the frequency placement of the signal.
We summarize the principal findings, assess the prospects for observational
detection, and identify the most promising directions for future work.

\subsection{Summary of main findings}

Our results  admit a
concise organizing principle: \emph{any modification to $T(M)$ that reduces
the temperature below the standard Hawking value shifts the GW spectral peak
to lower frequencies, while any modification that introduces a hard cutoff
sharpens the peak and steepens the high-frequency tail}.  The two effects
can operate independently or in combination, and together they span the full
range of qualitative behaviors seen in the figures.

More specifically, the main findings of this work are as follows.

\textit{Spectral diversity.}---The six modified-temperature models studied
here produce present-day GW spectra that differ from the standard Hawking
result by up to ten or more decades in peak frequency and by order-of-magnitude
differences in spectral width and shape.  Even models that share the same
qualitative behavior---such as the LQG and NC scenarios, both of which
predict a minimum mass and a vanishing temperature at $M_{\min}$---are
distinguishable through their spectral morphology, since the NC suppression
is quadratic in $M_{\min}/M$ while the LQG suppression is only square-root,
leading to measurably different peak widths and asymptotic tails at comparable
parameter values.

\textit{Parameter sensitivity.}---The location of the spectral peak is
exquisitely sensitive to the model-defining parameter.  In the plateau and
cooling models, shifting $M_c$ from a few $M_{\rm Pl}$ to $10^3$--$10^6\,
M_{\rm Pl}$ moves the peak by six or more decades, sweeping it from the
standard GHz--THz band down into the Hz--kHz range accessible to
ground-based interferometers.  In the tunneling model, the analogous shift
requires only a modest change in $\gamma$, with the stopping mass
$M_{\rm stop} = \sqrt{\gamma}$ providing a direct spectral ruler.  These
sensitivities imply that even a coarse frequency measurement of a GW peak
would constrain the model parameter to within a factor of a few.

\textit{Role of initial mass.}---The initial black-hole mass $M_0$ acts as
a global frequency dial, but its effect is non-trivially entangled with the
model-specific scale.  For models whose characteristic parameter is small
compared to $M_0$, increasing $M_0$ shifts all spectral features by a
common factor of order $M_0^3$ in the standard Hawking limit.  For models
whose cutoff scale is a significant fraction of $M_0$, the qualitative
character of the spectrum changes: the black hole spends a larger fraction
of its lifetime in the unmodified regime, producing a standard-Hawking-like
shoulder on the high-frequency side and a sharp modified cutoff at lower
frequencies.  Jointly varying $M_0$ and the model parameter is therefore
necessary for a complete survey of the signal landscape.

\textit{Cosmological history as an additional dial.}---The expansion history
between evaporation and today introduces an independent layer of model
dependence that is in principle separable from the black-hole physics.  An
early matter-dominated era reduces the redshift experienced by the emitted
gravitons, shifting all spectral features to higher present-day frequencies
and imprinting a characteristic $\Omega_{\rm GW}\propto f_0^{-2}$ break
at the reheating scale; kination does the opposite, pushing the spectrum
to lower frequencies and generating a blue $f_0^{+2}$ high-frequency tilt.
Crucially, these cosmological imprints are \emph{spectral-shape effects},
not merely rescalings of the peak frequency, so a broadband measurement can
in principle disentangle them from the black-hole temperature modifications
of primary interest.

\textit{Reheating dilution as an amplitude suppressor.}---Figure~\ref{fig:OmegaGW_modelsRH}
introduces a qualitatively new result not captured by the frequency-shift
analysis alone: when PBHs dominate the pre-reheating energy budget (the
large-$\beta$ regime), evaporation triggers an epoch of matter-dominated
expansion between formation at $t_{\rm form} \sim GM/(\gamma c^3)$ and
complete evaporation at $t_{\rm ev} \propto M_0^3\,t_{\rm Pl}$.  The
graviton energy fraction is diluted by a factor
\begin{equation}
  \mathcal{D}_{\rm RH}
  = \left(\frac{t_{\rm form}}{t_{\rm ev}}\right)^{2/3}
  \propto M_0^{-4/3},
  \label{eq:dilution_conc}
\end{equation}
which suppresses $\Omega_{\rm GW}\,h^2$ by several orders of magnitude
relative to the $\Delta N_{\rm eff}$ ceiling and grows rapidly with $M_0$.
For $M_0 = 10\,M_{\rm Pl}$ the suppression is modest ($\mathcal{D}_{\rm RH}
\sim 10^{-2}$), but for $M_0 = 1000\,M_{\rm Pl}$ it reaches $\sim 10^{-6}$--$10^{-7}$,
rendering the reheating-dominated signal undetectable with any instrument
concept considered here.  Crucially, the dilution is stronger for heavier
PBHs, so the large-$\beta$ scenario is simultaneously the most interesting
cosmologically---it connects PBH evaporation directly to reheating---and
the most challenging observationally.  The most favorable detection targets
within this scenario are therefore the lightest PBHs, $M_0 \sim
\mathcal{O}(10)\,M_{\rm Pl}$, for which $\mathcal{D}_{\rm RH}$ is least
severe and the spectral peak may still lie within a technologically
accessible frequency band.

\textit{Cosmology-independence of the relative spectral displacement.}---A
key implication of Fig.~\ref{fig:OmegaGW_modelsRH} is that, while the
\emph{absolute} position of the spectral peak depends strongly on the
assumed redshift prescription, the \emph{relative} displacement between
the standard Hawking peak and the modified-model peak is cosmology-independent:
both spectra are shifted by the same scale-factor ratio under any given
prescription.  Likewise, the amplitude suppression from reheating dilution
applies uniformly to both the standard and modified curves, preserving
their ratio.  This invariance means that the spectral-shape difference
and the ratio of peak frequencies---rather than their individual values---are
the cleanest observables for constraining $T(M)$, because they are
insensitive to the uncertain expansion history between evaporation and today.

\textit{Normalization and $N_{\rm eff}$ as a lever.}---Normalizing all
spectra to saturate $\Delta N_{\rm eff} \leq 0.3$ provides a
physically motivated and model-independent upper envelope.  A corollary of
this normalization is that models which confine their emission to a narrow
frequency band are necessarily brighter at the peak, making them
simultaneously the most spectrally distinctive and the easiest targets for
narrowband detectors.  A confirmed detection at or near the $\Delta N_{\rm eff}$
ceiling would immediately establish that Planck-scale black holes contribute
a cosmologically significant fraction of the relativistic energy budget at
BBN, a conclusion of profound importance independent of which specific
quantum-gravity model is responsible.  When reheating dilution is operative,
however, the observed amplitude can lie many orders of magnitude below this
ceiling; any such detection would then provide a direct measure of
$\mathcal{D}_{\rm RH}$ and hence of $\beta$, the initial PBH abundance.

\subsection{Limits on the observability of Planck-scale evaporation signatures}

Despite the encouraging breadth of the signal landscape, several fundamental
and practical limitations constrain the observability of the signatures
discussed in this work.

\textit{Frequency gap.}---The standard Hawking prediction for $M_0 \sim
10$--$100\,M_{\rm Pl}$ places the spectral peak in the GHz--THz regime,
far above the design bands of current and near-term interferometric detectors.
Only modified-temperature models with large characteristic scales---or
initial masses well above $100\,M_{\rm Pl}$---shift the peak into the kHz--MHz
range accessible to ET and CE.  The sub-kHz window, although in principle
reachable for extreme parameter choices, would require either very large
$M_0$ or modifications so dramatic that the signal becomes a marginal
perturbation on cosmological scales.  Absent a breakthrough in broadband
sensitivity above $\sim10^4\,\mathrm{Hz}$, the most generic predictions
of quantum-gravity--inspired evaporation scenarios will remain inaccessible
to ground-based interferometry alone.

\textit{Cosmological degeneracies and the three-prescription spread.}---As
illustrated explicitly in Fig.~\ref{fig:OmegaGW_modelsRH}, the same quantum-gravity
model at a fixed $M_0$ can populate any frequency window from $\lesssim10^5$~Hz
to $\gtrsim10^{15}$~Hz depending on the assumed cosmological history.
Extrapolating the radiation-dominated redshift formula to $T_{\rm Pl}$
(Prescription~1) places the peak at the lowest accessible frequencies;
evaluating the redshift at $T_{\rm RH,max}(t_{\rm ev})$ with the associated
matter-domination dilution (Prescription~2) shifts it upward by several
decades while simultaneously suppressing the amplitude; and treating the
PBHs as a subdominant component evaporating into a radiation-dominated
background (Prescription~3) yields the highest peak frequency at
undiminished amplitude.  These three curves span the plausible envelope of
cosmological uncertainty and should be regarded as bounding the expected
signal location rather than selecting a unique prediction.  Breaking the
resulting degeneracy between $T(M)$, $M_0$, and the redshift history
requires either independent external constraints on $T_{\rm RH}$---from,
e.g., thermally produced dark-matter abundances or CMB spectral distortions---or
a spectral-shape measurement precise enough to distinguish the morphological
signatures of different expansion histories.

\textit{Resonant detector requirements.}---The resonant-cavity sensitivity
curves overlaid on our figures represent aggressive but not implausible
benchmarks, requiring magnetic fields $B\sim25$--$35\,\mathrm{T}$, quality
factors $Q\sim10^{10}$--$10^{11}$, effective volumes $V\sim10^{1}$--$10^{2}\,
\mathrm{m}^3$, and system noise temperatures $T_{\rm sys}\sim\mathrm{few}
\times10\,\mathrm{mK}$ sustained over multi-year observation campaigns.
The futuristic UVHF concept shown in Fig.~\ref{fig:OmegaGW_modelsRH}---with
effective volume $V_{\rm eff} \sim 10^8\,\mathrm{m}^3$, $Q \sim 10^{14}$,
and $T_{\rm sys} \sim 1\,\mathrm{mK}$---is considerably more demanding and
should be understood as a long-range aspirational target rather than a
near-term projection.  The key technical tensions---maintaining ultra-high
$Q$ in strong magnetic fields, scaling to large volumes through cavity
arrays, and achieving quantum-limited readout over decade-long scans---are
identified explicitly in Sec.~\ref{sec:redshifting}, and none has been
resolved at the required level.  The projections presented here should
therefore be understood as targets for a new generation of instruments
rather than forecasts for existing technology.

\textit{Model dependence of the toy framework.}---The numerical spectra
presented in this work are based on a phenomenological, dimensionless
evaporation model that captures the qualitative physics of Hawking radiation
but does not include several quantitatively important effects: spin-dependent
greybody factors, the full Standard Model particle spectrum and its
energy-dependent emission rates, possible angular momentum of the initial
PBH population, and the detailed cosmological mass function of PBHs at
formation.  Incorporating these refinements will shift the spectral peak
and alter the shape at the factor-of-a-few level, and a fully quantitative
prediction for a specific model will require dedicated microphysical
calculations beyond the scope of this work.

\subsection{Future directions}

The results presented here motivate several concrete directions for
follow-up investigation.

\textit{Multi-messenger connections.}---The same evaporation event that
produces the GW background also generates a spectrum of Standard Model
particles.  The non-gravitational Hawking products---photons, neutrinos, if produced below the respective thermalization scale, as well as light beyond-Standard-Model states, if they exist---contribute to the
diffuse cosmic backgrounds at MeV--GeV energies and to the dark matter
abundance if stable Planck-scale remnants form~\cite{Chen:2015leg,
Carr:2020gox}.  Jointly analyzing the GW signal and these particle
backgrounds would provide correlated constraints on both $T(M)$ and the
initial PBH mass function, substantially reducing the degeneracies discussed
above.  In the PBH-domination scenario specifically, where
$\mathcal{D}_{\rm RH}$ suppresses the GW amplitude by a precisely
computable factor, the non-gravitational
particle yield provides an independent handle on $\beta$, thereby breaking
the degeneracy between the dilution-suppressed GW amplitude and a
genuinely sub-dominant PBH contribution.  In scenarios where PBH evaporation
drives reheating, additional connections to CMB spectral distortions, baryon
asymmetry generation, and the effective number of relativistic species at
BBN provide a rich multi-observable program~\cite{Papanikolaou:2020qtd,
Domenech:2021ztg}.

\textit{Quantum gravity phenomenology.}---Each of the temperature--mass
relations studied here was motivated by a specific quantum-gravity framework,
but the connection between the phenomenological parameters ($M_c$, $M_{\min}$,
$\gamma$, $M_H$) and the fundamental constants of those theories (e.g. the
GUP deformation parameter $\beta$, the LQG polymer scale, the string length)
has been left at the level of order-of-magnitude correspondence.  A tighter
mapping between microphysical parameters and spectral observables would
substantially increase the diagnostic power of a GW detection:
for instance, the string length inferred from a Hagedorn-shoulder measurement
could be compared directly with constraints from particle phenomenology
and collider experiments, providing a genuinely cross-disciplinary test of
quantum gravity.

\textit{Improved evaporation modeling.}---Several physically important
effects not included in the present framework warrant dedicated study.
The role of black-hole spin is particularly notable: rapidly rotating PBHs
emit a substantially larger fraction of their power in gravitons than
non-rotating ones, and they spin down through superradiant emission before
the final Planckian phase, potentially modifying both the overall GW
amplitude and the spectral shape~\cite{Page:1976df}.  A systematic study of
the Kerr case for each of the temperature models considered here would
determine whether spin introduces new spectral features---such as polarization
anisotropy or a distinctive high-frequency bump associated with the spindown
epoch---that could serve as additional discriminants.  Similarly, extending
the framework to include a distribution of initial masses (a PBH mass
function) rather than a single $M_0$ would smear the spectral peak and could
either wash out or accentuate model-specific features depending on the
width of the distribution.  The reheating dilution factor
$\mathcal{D}_{\rm RH}$ would also acquire a distribution in this case,
since $\mathcal{D}_{\rm RH} \propto M_0^{-4/3}$ depends sensitively on
mass; heavier members of the PBH population would contribute
negligibly to the GW background in the PBH-domination regime,
effectively imposing a light-mass selection on the detectable signal.

\textit{Refined treatment of the cosmological redshift history.}---The
three-prescription comparison shown in Fig.~\ref{fig:OmegaGW_modelsRH} motivates a more
systematic treatment of the pre-reheating expansion history as a
\emph{simultaneous} target of inference, rather than a source of uncertainty
to be marginalized over.  A Bayesian framework that jointly samples
$(M_0, \theta, T_{\rm RH}, \beta)$---where $\theta$ denotes the
model-specific quantum-gravity parameter---using the spectral shape,
peak position, and amplitude as observables could in principle separate these
contributions from a single broadband detection.  The spectral slopes
introduced by early matter domination ($\Omega_{\rm GW} \propto f_0^{-2}$)
and kination ($\Omega_{\rm GW} \propto f_0^{+2}$), together with the
amplitude suppression from reheating dilution, provide enough independent
constraints to make such a joint inference tractable if the
signal-to-noise ratio is sufficient across multiple frequency decades.

\textit{High-frequency detector development.}---The most direct implication
of this work for experiment is the identification of specific frequency--
amplitude targets that quantum-gravity--motivated GW spectra are expected
to occupy.  The resonant-cavity benchmarks adopted here span the MHz--GHz
regime and correspond to modifications with $M_c \sim 10^3$--$10^6\,M_{\rm Pl}$,
$M_{\min}/M_0 \sim 0.1$--$1$, or $\sqrt{\gamma}/M_0 \sim 0.1$--$1$---ranges
that are physically well motivated and not already excluded by any existing
constraint.  Even for Prescription~3
(the most favorable cosmological scenario) the signal touches the sensitivity
envelope of the GHz or MHz cavity concepts only in a narrow region.  A phased experimental program, beginning with
proof-of-concept GHz cavities and progressively extending to MHz-scale
lumped-element arrays and ultimately to UVHF resonant concepts in the THz
band, would systematically explore this landscape.  Crucially, the
spectral-shape information available from a broadband resonant array---as
opposed to a single narrowband detector---is essential for both model
discrimination and cosmological separation, making the case for investing
in frequency-agile or multiplexed detector architectures rather than
monolithic single-frequency instruments.

\textit{Theoretical completeness.}---Finally, this work has treated
the modified temperature--mass relations as independent phenomenological
inputs, without examining whether they are mutually consistent or whether
they can be embedded in a single ultraviolet-complete framework.  A more
ambitious program would seek a unified description of black-hole
thermodynamics that interpolates smoothly between the well-tested
semiclassical regime and the quantum-gravitational endpoint, making
predictions for all thermodynamic quantities---temperature, entropy,
heat capacity, and luminosity---from a single set of first-principles
assumptions.  The gravitational-wave spectrum, by virtue of its sensitivity
to the entire evaporation history rather than just the endpoint, is
uniquely suited to constrain such a unified description and thereby
discriminate between competing proposals for the ultraviolet completion
of gravity.

In summary, high-frequency gravitational waves from the evaporation of
near-Planck-mass primordial black holes represent a uniquely motivated
and potentially accessible window into quantum gravity.  Unlike most
proposed probes of Planck-scale physics, the Hawking evaporation process
has a well-understood semiclassical limit against which departures can
be systematically parameterized, and the gravitational-wave signal
it produces is, in principle, calculable within each quantum-gravity
framework to the extent that the effective temperature--mass relation is
specified.  Our analysis also highlights how the cosmological environment at the time of evaporation
is not merely a spectator, but an active participant that can shift the
spectral peak by many decades and suppress the amplitude by orders of
magnitude.  Separating the quantum-gravity signal from these cosmological
effects---and ultimately exploiting them as a complementary probe of the
pre-reheating universe---is both the principal challenge and the
principal opportunity of this observational program.  The observational
challenge is formidable, but the physical stakes---simultaneous empirical
access to the quantum structure of spacetime and the thermodynamic history
of the early universe---justify the ambition.  We hope that the framework
developed here will serve as a useful foundation for both theoretical
refinements and experimental planning as the high-frequency
gravitational-wave frontier opens in the coming decades.

\appendix
\section{Temperature–mass relations used in the toy study}\label{sec:app}

Note that everything is in dimensionless units, with masses in units of $M_{\rm Pl}$, $A=1/(8\pi)$.

\begin{equation}
T_{\rm std}(M)=\frac{A}{M}\,.
\end{equation}

\begin{equation}
T_{\rm plat}(M)=\frac{A}{\sqrt{M^{2}+M_{c}^{2}}}\,,
\qquad (M_c>0)\,.
\end{equation}

\begin{equation}
T_{\rm cool}(M)=A\,\frac{M}{M^{2}+M_{c}^{2}}\,,
\qquad (M_c>0)\,.
\end{equation}

\begin{equation}
T_{\rm GUP}(M)=A\,\frac{2M}{\beta}\left[\,1-\sqrt{1-\frac{\beta}{M^{2}}}\,\right]\,,
\qquad M\ge \sqrt{\beta}\,,
\end{equation}
where $\beta>0$ is the deformation parameter (and the square-root enforces a minimal mass scale).

\begin{equation}
T_{\rm LQG}(M)=\frac{A}{M}\sqrt{1-\left(\frac{M_{\min}}{M}\right)^{2}}\,,
\qquad M\ge M_{\min}\,.
\end{equation}

\begin{equation}
T_{\rm NC}(M)=\frac{A}{M}\left[\,1-\left(\frac{M_{\min}}{M}\right)^{2}\right]\,,
\qquad M\ge M_{\min}\,.
\end{equation}

\begin{equation}
T_{\rm tun}(M)=\frac{A}{M}\left(1-\frac{\gamma}{M^{2}}\right)\,,
\qquad M\ge \sqrt{\gamma}\,,
\end{equation}
with $\gamma>0$.

\begin{equation}
T_{\rm Hag}(M)=\frac{T_H}{1+\left(\frac{M}{M_H}\right)^{p}}\,,
\qquad T_H\equiv \frac{A}{M_H}\,,
\end{equation}
so that for $p=1$ one has $T_{\rm Hag}(M)\to A/M$ at $M\gg M_H$, while for $M\ll M_H$ the temperature saturates to $T_H$.

\bibliographystyle{apsrev4-2}
\bibliography{references}

\end{document}